\documentclass[aps,prd,twocolumn,showpacs]{revtex4-1}

\usepackage{wallpaper}
\usepackage{bm}
\usepackage{amsmath}
\usepackage[colorlinks,linkcolor=blue,anchorcolor=blue,citecolor=red]{hyperref}
\usepackage{CJK}

\begin{document}
\begin{CJK*}{GBK}{song}

\title{Interface effects of quark matter: Light-quark nuggets and compact stars}
\author{Cheng-Jun~Xia$^{1}$}
\email{cjxia@yzu.edu.cn}
\author{Jian-Feng Xu$^{2}$}
\email{jfxu@aynu.edu.cn}
\author{Guang-Xiong~Peng$^{3,4,5}$}
\email{gxpeng@ucas.ac.cn}
\author{Ren-Xin~Xu$^{6,7}$}
\email{r.x.xu@pku.edu.cn}

\affiliation{$^{1}${Center for Gravitation and Cosmology, College of Physical Science and Technology, Yangzhou University, Yangzhou 225009, China}
\\$^{2}${School of Physics \& Electrical Engineering, AnYang Normal University, AnYang 455000, China}
\\$^{3}${School of Nuclear Science and Technology, University of Chinese Academy of Sciences, Beijing 100049, China}
\\$^{4}${Theoretical Physics Center for Science Facilities, Institute of High Energy Physics, P.O. Box 918, Beijing 100049, China}
\\$^{5}${Synergetic Innovation Center for Quantum Effects and Application, Hunan Normal University, Changsha 410081, China}
\\$^{6}${School of Physics, Peking University, Beijing 100871, China}
\\$^{7}${Kavli Institute for Astronomy and Astrophysics, Peking University, Beijing 100871, China}}

\date{\today}

\begin{abstract}
The interface effects of quark matter play important roles in the properties of compact stars and small nuggets such as strangelets and $ud$QM nuggets. By introducing a density derivative term to the Lagrangian density and adopting Thomas-Fermi approximation, we find it is possible to reproduce the results obtained by solving Dirac equations. Adopting certain parameter sets, the energy per baryon of $ud$QM nuggets decreases with baryon number $A$ and become more stable than nuclei at $A\gtrsim 300$. The effects of quark matter symmetry energy are examined, where $ud$QM nuggets at $A\approx 1000$ can be more stable than others if large symmetry energy is adopted. In such cases, larger $ud$QM nuggets will decay via fission and the surface of an $ud$QM star will fragment into a crust made of $ud$QM nuggets and electrons, which resembles the cases of a strange star's crust. The corresponding microscopic structures are then investigated adopting spherical and cylindrical approximations for the Wigner-Seitz cells, where the droplet phase is found to be the most stable configuration with $ud$QM stars' crusts and $ud$QM dwarfs made of $ud$QM nuggets ($A\approx 1000$) and electrons. For the cases considered here, the crust thickness of $ud$QM stars is typically $\sim$200 m, which reaches a few kilometers if we neglect the interface effects and adopt Gibbs construction. The masses and radii of $ud$QM dwarfs are smaller than typical white dwarfs, which would increase if the interface effects are neglected.
\end{abstract}


\maketitle
\end{CJK*}

\section{\label{sec:intro}Introduction}
As we increase the density of baryonic matter, a deconfinement phase transition takes place and forms quark matter. The strange quark matter (SQM, comprised of $u$, $d$, and $s$ quarks) obtained in this process was expected to be the QCD ground state~\cite{Bodmer1971_PRD4-1601, Witten1984_PRD30-272, Terazaw1989_JPSJ58-3555}, where SQM objects with various sizes may exist in the Universe, e.g., strangelets~\cite{Farhi1984_PRD30-2379, Berger1987_PRC35-213, Gilson1993_PRL71-332, Peng2006_PLB633-314}, nuclearites~\cite{Rujula1984_Nature312-734, Lowder1991_NPB24-177}, meteorlike compact ultradense objects~\cite{Rafelski2013_PRL110-111102}, and strange stars~\cite{Itoh1970_PTP44-291, Alcock1986_ApJ310-261, Haensel1986_AA160-121}. However, in the framework of chiral models, SQM is unstable due to a too large $s$ quark mass~\cite{Buballa1999_PLB457-261, Klahn2015_ApJ810-134}. An interesting proposition was raised recently suggesting that, instead of SQM, the nonstrange quark matter ($ud$QM) may be the true ground state~\cite{Holdom2018_PRL120-222001}. This indicates the possible existence of $ud$QM nuggets and $ud$QM stars, while ordinary nuclei do not necessarily decay into $ud$QM nuggets due to a large enough surface tension~\cite{Holdom2018_PRL120-222001, Xia2020_PRD101-103031, Wang2021_Galaxies9-70}. Extensive investigations on the properties of nonstrange quark stars were then carried out in recent years~\cite{Zhao2019_PRD100-043018, Zhang2020_PRD101-043003, Cao2020, Zhang2021_PRD103-063018, Yuan2022}.

One important factor that affects the properties of strangelets and $ud$QM nuggets is the interface effects of quark matter, and in particular, the surface tension $\sigma$ that accounts for the energy contribution. Adopting the MIT bag model with a bag constant $B$, it was found that a small strangelet can be destabilized substantially at $\sigma^{1/3}\approx B^{1/4}$~\cite{Farhi1984_PRD30-2379}, while the minimum baryon number for metastable strangelets $A_\mathrm{min} \propto \sigma^3$~\cite{Berger1987_PRC35-213, Berger1989_PRD40-2128}. If the surface tension is smaller than a critical value with $\sigma<\sigma_\mathrm{crit}$, there exist strangelets at a certain size that are more stable than others~\cite{Heiselberg1993_PRD48-1418}. This leads to significant implications for SQM objects, where large strangelets are expected to decay via fission~\cite{Alford2006_PRD73-114016} and a strange star's surface fragments into a crystalline crust comprised of strangelets and electrons~\cite{Jaikumar2006_PRL96-041101}. For $ud$QM nuggets, due to the absence of $s$ quarks, the critical surface tension $\sigma_\mathrm{crit}$ is likely larger than that of strangelets~\cite{Alford2006_PRD73-114016, Jaikumar2006_PRL96-041101, Alford2008_PRC78-045802}. According to the recent estimations with various effective models~\cite{Oertel2008_PRD77-074015, Palhares2010_PRD82-125018, Pinto2012_PRC86-025203, Kroff2015_PRD91-025017, Garcia2013_PRC88-025207, Ke2014_PRD89-074041, Mintz2013_PRD87-036004, Gao2016_PRD94-094030, Xia2018_PRD98-034031, Fraga2019_PRD99-014046, Lugones2019_PRC99-035804}, it was shown that the surface tension value is likely small, i.e., $\sigma \lesssim 30\ \mathrm{MeV/fm}^{2}$. In such cases, it is relatively easy for $ud$QM to fulfill the condition $\sigma<\sigma_\mathrm{crit}$, indicating the possible existence of crusts inside $ud$QM quark stars and a new family of white dwarfs ($ud$QM dwarfs). Additionally, it was shown that quark matter may have a large symmetry energy~\cite{Chu2014_ApJ780-135, Jeong2016_NPA945-21, Chen2017_NPR34-20, Chu2019_PRC99-035802, Wu2019_AIPCP2127-020032}, which further increases $\sigma_\mathrm{crit}$ and affects the properties of $ud$QM objects~\cite{Wang2021_Galaxies9-70}. Note that if the strangelets or $ud$QM nuggets inside those stars are small, the finite size effects are dominated by curvature effects rather than by surface tension~\cite{Lugones2021_PRC103-035813}. In such cases, the criterion $\sigma<\sigma_\mathrm{crit}$ is no longer valid and the contribution of curvature effects needs to be accounted for. Meanwhile, the prediction of a small $\sigma$ is neither unanimous nor conclusive. The surface tension value may be strongly enhanced by new terms arising from vector interactions, which would inhibit the formation of quark star crusts~\cite{Lugones2013_PRC88-045803, Lugones2021_PRD104-L101301}.

In addition to the surface tension, other aspects of the interface effects also play important roles. Due to confinement of color charge, the quark wave functions approach to zero on the surface, which leads to quark depletion and contributes to the interface effects. For strangelets, it was shown that quark depletion has sizable impacts on their sizes and charge properties~\cite{Xia2016_SciBull61-172, Xia2016_SciSinPMA46-012021_E, Xia2016_PRD93-085025, Xia2017_JPCS861-012022, Xia2017_NPB916-669}, while the curvature term plays an important role especially for small objects~\cite{Madsen1993_PRL70-391, Madsen1993_PRD47-5156, Madsen1994_PRD50-3328, Lugones2021_PRC103-035813}. To account for those effects, it is favorable to solve the Dirac equations and obtain the structure of quark-vacuum interface self-consistently, e.g., those in Refs.~\cite{Oertel2008_PRD77-074015, Xia2018_PRD98-034031}. Nevertheless, for large objects, to obtain all the wavefunctions of quarks requires extensive computational resource, which limits its application in various scenarios. It is thus necessary to adopt a simpler approach that accounts for the general properties of interface effects. This was done in the framework of bag model, where an analytical formula that introduces a modification to the density of states was proposed, i.e., the multiple reflection expansion (MRE) method~\cite{Berger1987_PRC35-213, Berger1991_PRC44-566, Madsen1993_PRL70-391, Madsen1993_PRD47-5156, Madsen1994_PRD50-3328}. However, the MRE method describes the depleted quarks with a delta function, which oversimplifies the surface structure and is not applicable for other models with more realistic confining potential, e.g., the linear confinement in the framework of equivparticle model~\cite{Xia2018_PRD98-034031}. In such cases, we consider an effective approach that introduces a density derivative term to the Lagrangian density, which well reproduces the surface and curvature contributions with a detailed surface structure.

In this work we thus examine the effectiveness of the density derivative term in describing the interface effects, where the properties of strangelets and $ud$QM nuggets are investigated. For $ud$QM nuggets that are more stable at certain sizes, we further investigate the nonuniform structures comprised of those nuggets and electrons, which may persist inside quark stars and white dwarfs~\cite{Wang2021_Galaxies9-70}. It is found that for a given parameter set, a quark star with $M\gtrsim 0.03 M_{\odot}$ can support a stable crust ($\sim$200 m thick) on its surface, while the maximum mass of $ud$QM dwarfs decreases and becomes smaller than $1.4 M_{\odot}$ if the interface effects are accounted for.
The paper is organized as follows. In Sec.~\ref{sec:the} we present the theoretical framework of the equivparticle model, where a density derivative term was added to the original Lagrangian density to mimic the interface effects. To investigate the impact of symmetry energy, an isospin dependent term is added to the quark mass scaling. Then the structures of strangelets, $ud$QM nuggets, and the crystalline structures of quark star crusts are obtained adopting Thomas-Fermi approximation. The numerical results are presented in Sec.~\ref{sec:num}. We draw our conclusion in Sec.~\ref{sec:con}.

\section{\label{sec:the}Theoretical framework}
To investigate the properties of quark matter and the corresponding objects, we adopt the equivparticle model that treats the strong interaction with density-dependent quark masses~\cite{Peng2000_PRC62-025801, Wen2005_PRC72-015204, Wen2007_JPG34-1697, Xia2014_SCPMA57-1304, Chen2012_CPC36-947, Chang2013_SCPMA56-1730, Xia2014_PRD89-105027, Chu2014_ApJ780-135, Hou2015_CPC39-015101, Peng2016_NST27-98, Chu2017_PRD96-083019}. The Lagrangian density for equivparticle model is fixed by
\begin{eqnarray}
\mathcal{L} &=&  \sum_{i} \bar{\psi}_i \left[ i \gamma^\mu \partial_\mu - m_i(n_u, n_d, n_s) - e q_i \gamma^\mu A_\mu \right]\psi_i \nonumber\\
            &&{} - \frac{1}{4} A_{\mu\nu}A^{\mu\nu} + \mathcal{L}_\mathrm{der}+ \mathcal{L}_e+ \mathcal{L}_\mu,  \label{eq:Lgrg_all}\\
\mathcal{L}_\mathrm{der} &=& - \sum_{i,j} \frac{1}{2} \delta_V \partial_\mu  \left(\bar{\psi}_i \gamma_\nu \psi_i\right) \partial^\mu  \left(\bar{\psi}_j \gamma^\nu \psi_j\right), \label{eq:Lgrg_der} \\
\mathcal{L}_{e,\mu} &=& \bar{\psi}_{e,\mu} \left[ i \gamma^\mu \partial_\mu - m_{e,\mu} + e^2 \gamma^\mu A_\mu \right]\psi_{e,\mu},
\end{eqnarray}
where $\psi_i$ represents the Dirac spinor of quark flavor $i$, $m_i(n_\mathrm{b})$ the equivalent quark mass, $n_i=  \langle \bar{\psi}_i \gamma^0 \psi_i\rangle$ the number density, $q_i$ the charge ($q_u=2e/3$ and $q_d=q_s=-e/3$), and $A_\mu$ the photon field with the field tensor
\begin{equation}
A_{\mu\nu} = \partial_\mu A_\nu - \partial_\nu A_\mu.
\end{equation}
An additional density derivative term $\mathcal{L}_\mathrm{der}$ is introduced to account for the average interface effects, which was not included in our previous studies~\cite{Xia2018_PRD98-034031}. As will be illustrated in this work, by properly adjusting the effective parameter $\delta_V$ and adopting Thomas-Fermi approximation (TFA), we can reproduce the results obtained by solving the Dirac equation under mean-field approximation (MFA)~\cite{Xia2018_PRD98-034031, Xia2019_AIPCP2127-020029}. Note that if other contributions to the interface effects are considered~\cite{Oertel2008_PRD77-074015, Buballa2015_PPNP81-39}, the parameter $\delta_V$ may be altered. The Lagrangian density $\mathcal{L}_{e,\mu}$ accounts for the contribution of leptons ($e^-$, $\mu^-$) with $m_{e,\mu}$ being their masses, which should be included if electrons and/or muons are present.

In the past decades, various types of quark mass scalings were proposed to describe the strong interaction among quarks. For example, for density dependent quark masses
\begin{equation}
  m_i(n_{\mathrm b}) = m_{i0} + m_\mathrm{I}(n_{\mathrm b}),
\end{equation}
bag model suggests an inversely linear scaling $m_\mathrm{I} =  {B}/{3 n_\mathrm{b}}$~\cite{Fowler1981_ZPC9-271}, where $n_\mathrm{b}= \sum_{i=u,d,s}n_i$ is the baryon number density and $m_{u0}=2.2$ MeV, $m_{d0}=4.7$ MeV, $m_{s0}=96.0$ MeV the current masses of quarks~\cite{PDG2016_CPC40-100001}. An inversely cubic scaling $m_\mathrm{I} =  {D}{n_\mathrm{b}^{-1/3}}$ was derived if the contributions of linear confinement and in-medium chiral condensates were considered~\cite{Peng1999_PRC61-015201}. Further consideration of one-gluon-exchange interaction suggests $m_\mathrm{I} =   {D}{n_\mathrm{b}^{-1/3}} - C n_\mathrm{b}^{1/3}$~\cite{Chen2012_CPC36-947}, while perturbation theory at ultrahigh densities gives $m_\mathrm{I}= {D}{n_\mathrm{b}^{-1/3}}+C n_\mathrm{b}^{1/3}$~\cite{Xia2014_PRD89-105027}. An isospin dependent term was also introduced to examine the impacts of quark matter symmetry energy, which was given by $m_i(n_{\mathrm b}, \delta) = m_{i0} +  {D}{n_\mathrm{b}^{-1/3}} - \tau_i \delta D_I n_\mathrm{b}^{\alpha} \mbox{e}^{-\beta n_\mathrm{b}}$ with $\tau_i$ being the third component of isospin for quark flavor $i$ and $\delta = 3(n_d-n_u)/(n_d+n_u)$ the isospin asymmetry~\cite{Chu2014_ApJ780-135}. Recently, we have proposed a similar mass scaling~\cite{Wang2021_Galaxies9-70}
\begin{equation}
m_\mathrm{I}(n_{\mathrm b}, \delta) = {D}{n_\mathrm{b}^{-1/3}}+C n_\mathrm{b}^{1/3} + C_I \delta^2 n_\mathrm{b},  \label{Eq:mI}
\end{equation}
where the first term corresponds to linear confinement with a confinement parameter $D$~\cite{Peng1999_PRC61-015201}. Depending on the sign of $C$, the second term represents the contribution of one-gluon-exchange interaction ($C<0$)~\cite{Chen2012_CPC36-947} or the leading-order perturbative interaction ($C>0$)~\cite{Xia2014_PRD89-105027}. The third term accounts for the quark matter symmetry energy~\cite{Wang2021_Galaxies9-70}. In contrast to the mass scaling proposed in Ref.~\cite{Chu2014_ApJ780-135}, here $m_\mathrm{I}$ is identical for different quark flavor, i.e., neglecting the isovector-scalar channel.

Adopting the mass scaling in Eq.~(\ref{Eq:mI}), the mean field vector potential for quark type $i$ in MFA can be fixed according to variational method, which is obtained with
\begin{equation}
 V_i = \frac{\mbox{d} m_\mathrm{I}}{\mbox{d} n_i}\sum_{i=u,d,s}  n_i^\mathrm{s} + e q_i A_0  + 3\delta_V \nabla^2 n_\mathrm{b}. \label{Eq:Vi}
\end{equation}
Here $n_i^\mathrm{s} = \langle \bar{\psi}_i \psi_i \rangle$ represents the scalar density of quark flavor $i$. The density derivative term of quark masses in the right hand side is introduced to maintain thermodynamic self-consistency. In fact, for any effective models with density dependent masses or coupling constants, the density derivative terms appear naturally according to fundamental thermodynamics, which have been discussed extensively in various publications, e.g., those in Ref.~\cite{Brown1991_PRL66-2720, Lenske1995_PLB345-355, Wang2000_PRC62-015204, Peng2000_PRC62-025801, Torres2013_EPL101-42003, Dexheimer2013_EPJC73-2569, Xia2014_PRD89-105027}. The second term represents the contribution from Coulomb interaction and is fixed by
\begin{equation}
  \nabla^2 A_0 =- \sum_{i = u, d, s, e, \mu} q_i n_i. \label{Eq:A0}
\end{equation}
Note that the third term in Eq.~(\ref{Eq:mI}) has lead to the isovector contributions in the vector potentials of quarks. Since leptons have nothing to do with strong interaction and take constant masses, the first and third terms vanish and Eq.~(\ref{Eq:Vi}) becomes $V_{e,\mu} = - e^2 A_0$.

The total particle number $N_i$ is obtained by integrating the density $n_i(\vec{r})$, i.e.,
\begin{equation}
 N_i = \int n_i(\vec{r}) \mbox{d}^3 r. \label{Eq:axi}
\end{equation}
The total energy of the system is determined by
\begin{equation}
M= E_0 + \int \left[\frac{9}{2}\delta_V n_\mathrm{b}(\vec{r}) \nabla^2 n_\mathrm{b}(\vec{r})
     + \frac{1}{2}(\nabla A_0)^2\right] \mbox{d}^3 r. \label{eq:mass}
\end{equation}
Here $E_0$ represents the contributions of kinetic energy and strong interaction with density-dependent quark masses, which will be fixed based on different approaches introduced in the following.

\subsection{\label{sec:the_spher} Spherically symmetric objects}
For a spherically symmetric system, the Dirac spinor of fermions can be expanded as
\begin{equation}
 \psi_{n\kappa m}({\bm r}) =\frac{1}{r}
 \left(\begin{array}{c}
   iG_{n\kappa}(r) \\
    F_{n\kappa}(r) {\bm\sigma}\cdot{\hat{\bm r}} \\
 \end{array}\right) Y_{jm}^l(\theta,\phi)\:,
\label{EQ:RWF}
\end{equation}
with $G_{n\kappa}(r)/r$ and $F_{n\kappa}(r)/r$ being the radial wave functions for the upper and lower components, while $Y_{jm}^l(\theta,\phi)$ is the spinor spherical harmonics, i.e.,
\begin{equation}
 Y_{jm}^l = \sum_{l_z, s_z} \langle l,l_z; {1}/{2}, s_z | j, m \rangle Y_{l l_z}\chi_{1/2}^{s_z}.
\end{equation}
The quantum number $\kappa$ is defined by the angular momenta $(l,j)$ as $\kappa=(-1)^{j+l+1/2}(j+1/2)$. The Dirac equation for the radial wave functions in MFA is then obtained as
\begin{equation}
 \left(\begin{array}{cc}
  V_i + m_i                                            & {\displaystyle -\frac{\mbox{d}}{\mbox{d}r} + \frac{\kappa}{r}}\\
  {\displaystyle \frac{\mbox{d}}{\mbox{d}r}+\frac{\kappa}{r}} & V_i - m_i                      \\
 \end{array}\right)
 \left(\begin{array}{c}
  G_{n\kappa} \\
  F_{n\kappa} \\
 \end{array}\right)
 = \varepsilon_{n\kappa}
 \left(\begin{array}{c}
  G_{n\kappa} \\
  F_{n\kappa} \\
 \end{array}\right) \:.
\label{Eq:RDirac}
\end{equation}
For given radial wave functions, the scalar and vector densities can be determined by
\begin{subequations}
\begin{eqnarray}
 n_i^\mathrm{s}(r) &=&\frac{1}{4\pi r^2}\sum_{k=1}^{N_i}
 \left[|G_{k i}(r)|^2-|F_{k i}(r)|^2\right] \:, \\
 n_i(r) &=& \frac{1}{4\pi r^2}\sum_{k=1}^{N_i}
 \left[|G_{k i}(r)|^2+|F_{k i}(r)|^2\right] \:.
\end{eqnarray}%
\label{Eq:Density}%
\end{subequations}%
The total energy of the system can then be fixed by Eq.~(\ref{eq:mass}) with
\begin{eqnarray}
E_0 &=& \sum_{i}\left[\sum_{k=1}^{N_i} \varepsilon_{ki} - \int 4\pi r^2 n_i(r) V_i(r) \mbox{d}r \right].
\label{Eq:E0}
\end{eqnarray}
For strangelets and $ud$QM nuggets, we neglect the contribution of leptons. The Dirac Eq.~(\ref{Eq:RDirac}), mean field potentials in Eqs.~(\ref{Eq:mI}) and (\ref{Eq:Vi}), and densities in Eq.~(\ref{Eq:Density}) are solved via iteration in coordinate space with the grid width $0.005$ fm, where quarks occupy the lowest energy levels and reaches $\beta$-stability. More detailed discussion on the numerical treatments can be found in our previous studies~\cite{Xia2018_PRD98-034031, Xia2019_AIPCP2127-020029}.

\subsection{\label{sec:the_TFA} Thomas-Fermi approximation}
For a system with large number of particles, instead of solving the Dirac equation as indicated in Sec.~\ref{sec:the_spher}, the Thomas-Fermi approximation also provides relative good estimations for its properties. In the framework of TFA, we have $E_0=\int \mathcal{E}_0(\vec{r}) \mbox{d}^3 r$ for the total energy in Eq.~(\ref{eq:mass}), where the local kinetic energy density $\mathcal{E}_0(\vec{r})$ is fixed by
\begin{equation}
\mathcal{E}_0 = \sum_{i} \int_0^{\nu_i} \frac{d_i p^2}{2\pi^2} \sqrt{p^2+{m_i}^2}\mbox{d}p
              = \sum_{i} \frac {d_i{m_i}^4}{16\pi^{2}}f\left(\frac{\nu_i}{m_i}\right).
\end{equation}
Here $f(x) = \left[x(2x^2+1)\sqrt{x^2+1}-\mathrm{arcsh}(x) \right]$, $d_i$ (= 6 for quarks and 2 for leptons) the degeneracy factor, and $\nu_i$ the Fermi momentum of particle type $i$. The scalar and vector densities are determined by
\begin{equation}
n_i^\mathrm{s} =   \frac{d_i{m_i}^3}{4\pi^2} g\left(\frac{\nu_i}{m_i}\right),\ \
n_i  = \frac{d_i\nu_i^3}{6\pi^2},
\label{eq:dens_TFA}
\end{equation}
where $g(x) = x \sqrt{x^2+1} - \mathrm{arcsh}(x)$. Note that $m_i$, $n_{s}$, $n_i$, and $\mathcal{E}_0$ represent the local properties of quark matter and vary with the space coordinates, which can be determined by the constancy of chemical potentials in order for the system to be stable, i.e.,
\begin{equation}
\mu_i(\vec{r}) = \sqrt{{\nu_i(\vec{r})}^2+{m_i(\vec{r})}^2} + V_i(\vec{r}) = \rm{constant}. \label{eq:chem_cons}
\end{equation}
Once the density profiles satisfy Eq.~(\ref{eq:chem_cons}), the total energy $M$ of the system reaches a minimum. We thus adopt the imaginary time step method~\cite{Levit1984_PLB139-147} and readjust the density profiles iteratively. As convergency is reached with a vanishing deviation of local chemical potentials $\sum_{i} \langle \Delta \mu_i^2 \rangle$, the total mass of the system can be fixed by Eq.~(\ref{eq:mass}), where the equivalent mass and vector potentials are obtained with Eqs.~(\ref{Eq:mI}) and (\ref{Eq:Vi}).

Here we apply TFA to two types of objects, i.e., the spherically symmetric objects in the absence of leptons (strangelets and $ud$QM nuggets) and the Wigner-Seitz (WS) cell that describes the microscopic structures in a quark stars' crusts. Since quark propagation around the quark-vacuum interface is absent in TFA, it is essential to adopt the density derivative term $\mathcal{L}_\mathrm{der}$ in Eq.~(\ref{eq:Lgrg_der}) for finite-sized objects. The corresponding effective parameter $\delta_V$ is fixed by reproducing the results obtained by solving the wavefunctions as illustrated in Sec.~\ref{sec:the_spher}. For those predicting $ud$QM nuggets that are more stable than heavier ones, we then investigate the possible formation of crystalline structures in quark stars~\cite{Jaikumar2006_PRL96-041101, Alford2008_PRC78-045802}, where spherical and cylindrical approximations for the WS cell are adopted~\cite{Pethick1998_PLB427-7, Oyamatsu1993_NPA561-431, Maruyama2005_PRC72-015802, Togashi2017_NPA961-78, Shen2011_ApJ197-20}. In particular, the Laplace operator in Eqs.~(\ref{Eq:Vi}) and (\ref{Eq:A0}) are reduced to one-dimensional, i.e.,
\begin{eqnarray}
 \mathrm{1D:}\ \ \ \  && \nabla^2 \phi(\vec{r}) = \frac{\mbox{d}^2\phi(r)}{\mbox{d}r^2}, \label{eq:dif_1D} \\
 \mathrm{2D:}\ \ \ \  && \nabla^2 \phi(\vec{r}) = \frac{\mbox{d}^2\phi(r)}{\mbox{d}r^2} + \frac{1}{r} \frac{\mbox{d}\phi(r)}{\mbox{d}r}, \label{eq:dif_2D}\\
 \mathrm{3D:}\ \ \ \  && \nabla^2 \phi(\vec{r}) = \frac{\mbox{d}^2\phi(r)}{\mbox{d}r^2} + \frac{2}{r} \frac{\mbox{d}\phi(r)}{\mbox{d}r}, \label{eq:dif_3D}
\end{eqnarray}
with $\phi=n_\mathrm{b}$ and $A_0$. The density profiles and mean fields are then obtained with fast cosine transformation, where the reflective boundary conditions at $r=0$ and $r=R_\mathrm{W}$ are fulfilled with $R_\mathrm{W}$ being the WS cell radius. At fixed average baryon number density $n_\mathrm{b}$, we then search for the optimum cell size $R_\mathrm{W}$ by minimizing the energy per baryon, where the $\beta$-stability condition $\mu_u+\mu_e=\mu_u+\mu_\mu=\mu_d=\mu_s$ and global charge neutrality condition $\sum_iq_iN_i=0$ are satisfied. By assuming various dimensions with geometrical symmetries, five types of structures can be obtained based on TFA, i.e., the slab phase in Eq.~(\ref{eq:dif_1D}),  the rod/tube phases in Eq.~(\ref{eq:dif_2D}), and the droplet/bubble phases in Eq.~(\ref{eq:dif_3D}). A more detailed illustration on the numerical recipe can be found in Ref.~\cite{Xia2021_PRC103-055812, Xia2022_PRC105-045803}. Note that at small $n_\mathrm{b}$ electrons take up most of the space in a WS cell, leading to a too large cell size $R_\mathrm{W}$.  In such cases, as was done in Ref.~\cite{Xia2022_PRC105-045803}, we divide a WS cell into a core ($r<R_\mathrm{in}$) and a spherical shell ($R_\mathrm{in}<r\leq R_\mathrm{W}$). For the droplet phase, we take $R_\mathrm{in}=60$ fm.

\section{\label{sec:num}Results and discussions}

\subsection{\label{sec:num_slets}Strangelets}
\begin{table}
\caption{\label{tab:dltV}The coefficient $\delta_V$ for the density derivative term $\mathcal{L}_\mathrm{der}$ in Eq.~(\ref{eq:Lgrg_all}), where $n_0$ represents the density of SQM at $P=0$. The strength $C_I$ for the symmetry energy term in Eq.~(\ref{Eq:mI}) has little impact on the values of $\delta_\mathrm{v}$, which is taken as $C_I=0$ here.}
\begin{tabular}{cc|c|c} \hline \hline
 $C$    & $\sqrt{D}$ &   $n_0$    &  $\delta_\mathrm{v}$   \\
        &    MeV     & fm$^{-3}$  &  $10^{-10}$ MeV$^{4}$  \\   \hline
 $-0.5$ &  180       & 0.37       &  $-0.5 $    \\
   0    &  156       & 0.24       &  $-1.15$    \\
   0.1  &  150       & 0.16       &  $-1.5$     \\
   0.7  &  129       & 0.099      &  $-3.6 $    \\   \hline
\end{tabular}
\end{table}

To reproduce the results obtained by solving Dirac equations ($\delta_V=0$)~\cite{Xia2018_PRD98-034031, Xia2019_AIPCP2127-020029}, one needs to adjust the effective parameter $\delta_V$ for the density derivative term $\mathcal{L}_\mathrm{der}$ in Eq.~(\ref{eq:Lgrg_der}). Particularly, we reproduce the energy per baryon of $\beta$-stable ($\mu_u = \mu_d = \mu_s$) strangelets with TFA, where in Table~\ref{tab:dltV} the corresponding values of $\delta_V$ for various sets of parameters ($C$, $C_I=0$, $D$) are presented. It is found that $\delta_V$ takes negative values and its absolute value is increasing with the density $n_0$ at vanishing external pressure, which resembles our previous findings with both surface tension and curvature term increasing with $n_0$~\cite{Xia2018_PRD98-034031, Xia2019_AIPCP2127-020029}.

\begin{figure}
\includegraphics[width=\linewidth]{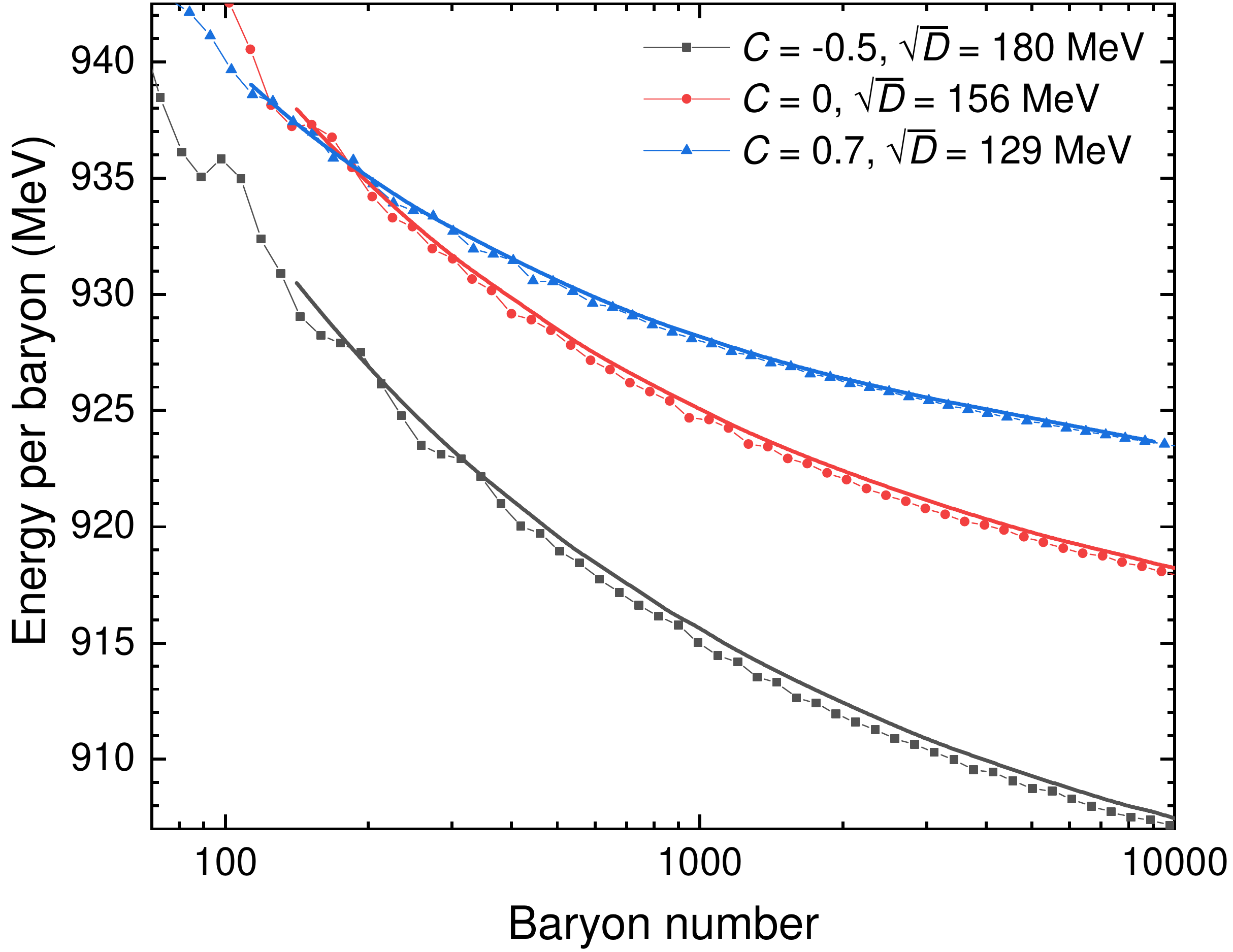}
\caption{\label{Fig:EpA_Slet} Energy per baryon ($M/A$) of $\beta$-stable strangelets obtained with TFA (solid line), which are compared with those predicted by explicitly solving Dirac equations (symbol+line).}
\end{figure}

Few examples on the obtained energy per baryon of $\beta$-stable strangelets are presented in Fig.~\ref{Fig:EpA_Slet}, which are decreasing with baryon number $A=(N_u+N_d+N_s)/3$ due to the interface effects. Adopting the values of $\delta_\mathrm{v}$ indicated in Table~\ref{tab:dltV}, it is shown that our calculation with TFA well reproduces the results obtained by solving the Dirac equations of quarks. Note that the energy per baryon obtained with TFA varies smoothly with $A$. However, if we solve the Dirac equations and examine the energy per baryon of strangelets, there are slight fluctuations due to the shell effects in single particle levels of quarks. This causes deviations at small $A$, which are expected to become insignificant at large baryon numbers, e.g., $A\gtrsim 300$.

\begin{figure}
\includegraphics[width=\linewidth]{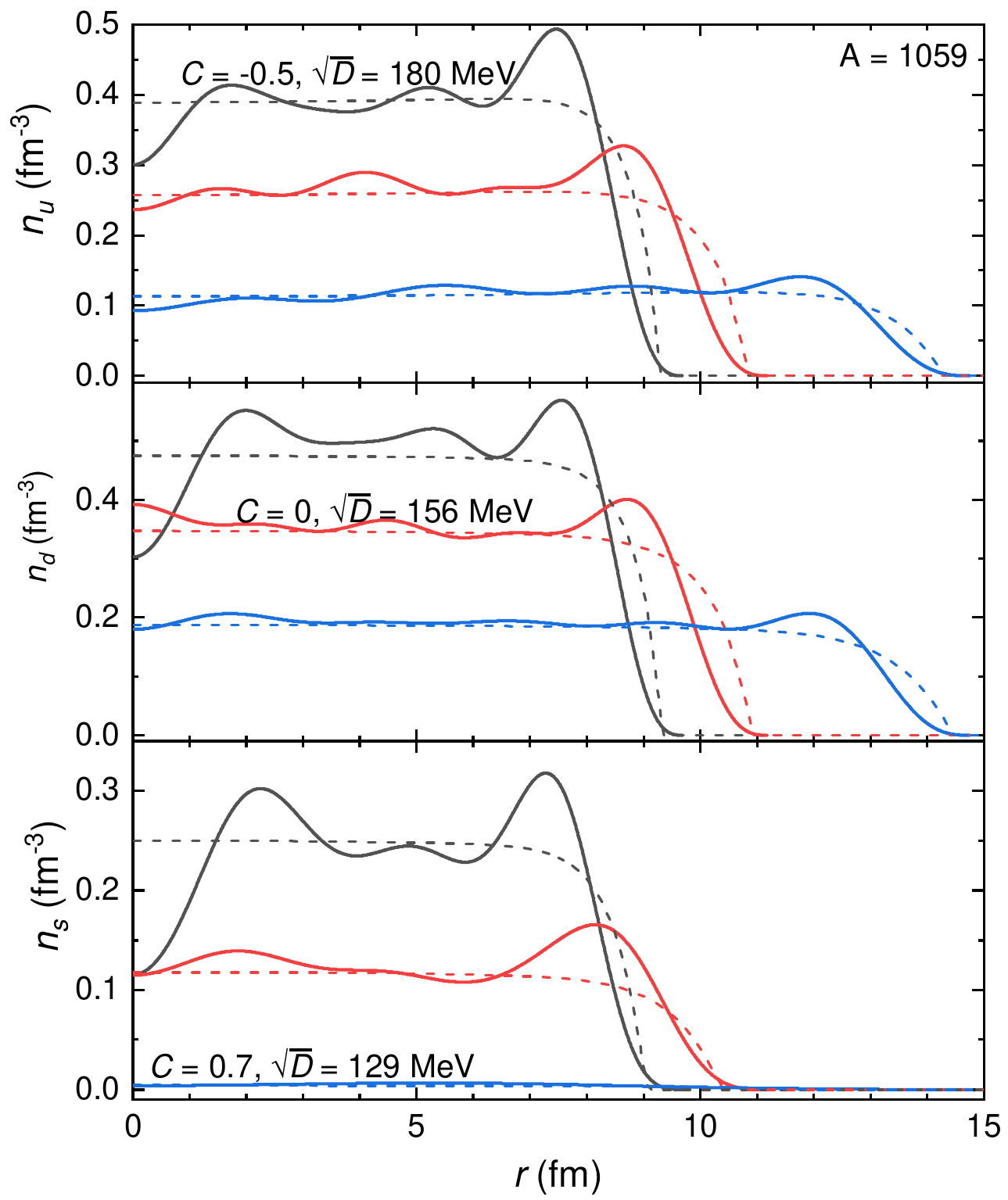}
\caption{\label{Fig:ni_Comp} Comparison between the density profiles in strangelets determined by wavefunctions with Eq.~(\ref{Eq:Density}) (solid) and TFA with Eq.~(\ref{eq:dens_TFA}) (dashed).}
\end{figure}

In the framework of TFA, the density profiles of strangelets at fixed baryon number $A$ are obtained by fulfilling the constancy of chemical potentials in Eq.~(\ref{eq:chem_cons}), where the local densities are fixed by Eq.~(\ref{eq:dens_TFA}). In Fig.~\ref{Fig:ni_Comp} we present the density profiles of strangelets at $A = 1059$. With the inclusion of the density derivative term $\mathcal{L}_\mathrm{der}$ in Eq.~(\ref{eq:Lgrg_der}), the density profiles obtained with TFA generally coincide with those from quark wave functions. Since there exist several nodes in quark wave functions, as indicated in Fig.~\ref{Fig:ni_Comp}, the density profiles obtained with Eq.~(\ref{Eq:Density}) fluctuate and do not take constant values even at the regions far from the surface, leading to minor discrepancies between those obtained with the two methods. It is expected that the discrepancies should eventually vanish at large enough $A$, where the TFA is expected to give a good description of the system. Nevertheless, in the surface region of a strangelet, the densities obtained with TFA are larger than those predicted by quark wave functions, where the situation reverses at regions slightly beneath the surface. Such a deviation on the surface is caused by the quark-vacuum interface, which can not be fixed by increasing the baryon number $A$.

\begin{figure}
\includegraphics[width=\linewidth]{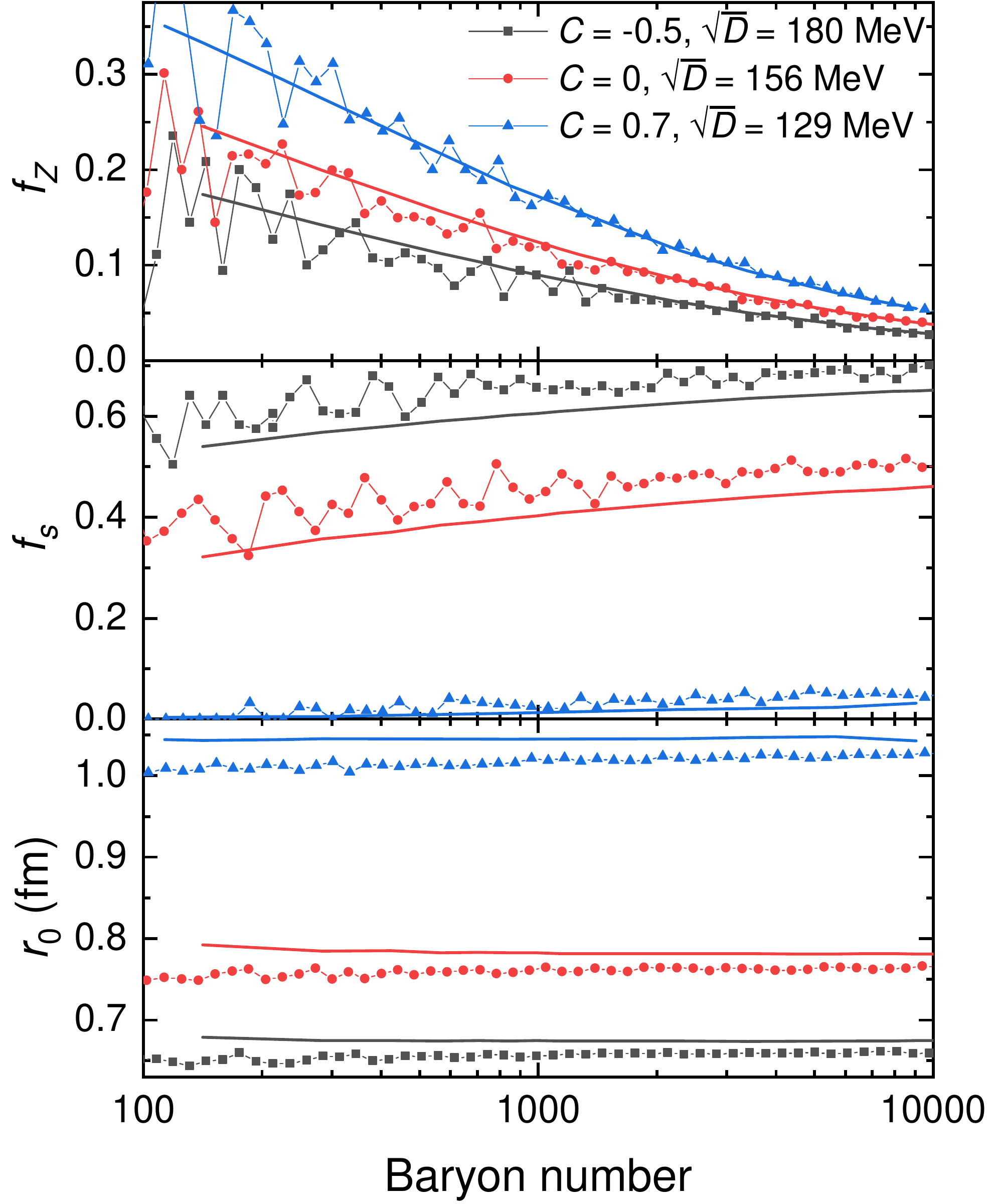}
\caption{\label{Fig:frac_Slet} Charge-to-mass ratio ($f_Z$), strangeness per baryon ($f_S$), and the ratio of root-mean-square radius to baryon number ($r_0$) for $\beta$-stable strangelets, in correspondence to those in Fig.~\ref{Fig:EpA_Slet}.}
\end{figure}

The charge-to-mass ratio ($f_Z = Z/A$), strangeness per baryon ($f_S = S/A$), and ratio of root-mean-square radius to baryon number ($r_0={\langle r^2 \rangle}^{1/2}/A^{1/3}$) for $\beta$-stable strangelets are presented in Fig.~\ref{Fig:frac_Slet}. The charge number and strangeness of a strangelet are fixed by $Z=(2N_u-N_d-N_s)/3$ and $S=N_s$ with $N_i$ being the total particle number of quark flavor $i$. Since the single particle levels of quarks are discrete, the sequential occupation of lowest energy levels causes fluctuations in the charge-to-mass ratio and strangeness per baryon, which will eventually become insignificant at large $A$. In such cases, as indicated in Fig.~\ref{Fig:frac_Slet}, the TFA with density derivative term well reproduces the results determined by solving Dirac equations. Nevertheless, there exist a slight deviation for both $f_S$ and $r_0$, where the TFA underestimates $f_S$ and overestimates $r_0$, which are mainly caused by the discrepancies of density profiles in the vicinity of quark-vacuum interface as indicated in Fig.~\ref{Fig:ni_Comp}.

\subsection{$ud$QM nuggets}
Now let us examine the properties of $ud$QM nuggets adopting the same parameter sets as strangelets in Sec.~\ref{sec:num_slets}. It is worth mentioning that in order to reproduce the experimental baryon masses with various effective models, the typical mass for strange quarks is usually $\sim$2-3 $m_{s0}$, e.g., the MIT bag model~\cite{DeGrand1975_PRD12-2060, Bernotas2004_NPA741-179, Maezawa2005_PTP114-317}, nonrelativistic quark cluster model~\cite{Oka1983_PLB130-365, Straub1988_PLB200-241}, and potential models~\cite{Jena2000_PRD63-014011}. In such cases, strangelets may not be stable due to the large strange quark masses, where $s$ quarks never emerge if the $\beta$-stability condition is fulfilled and strangelets are then converted into $ud$QM nuggets.

\begin{figure}
\includegraphics[width=\linewidth]{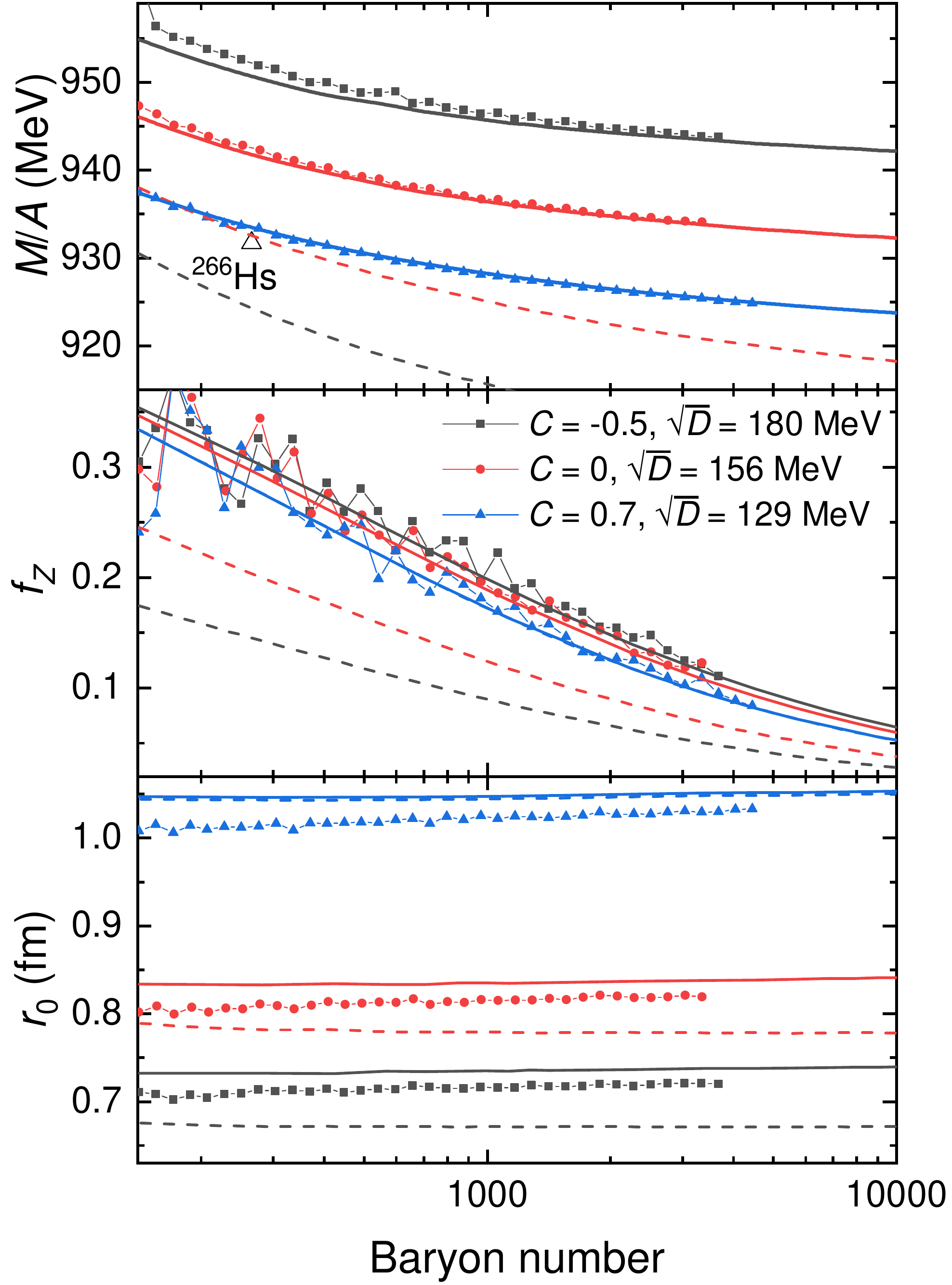}
\caption{\label{Fig:udQM_Efrac} Energy per baryon, charge-to-mass ratio, and ratio of root-mean-square radius to baryon number of $ud$QM nuggets. The dashed curves correspond to the properties of strangelets indicated in Figs.~\ref{Fig:EpA_Slet} and \ref{Fig:frac_Slet}, where the same parameter sets are adopted as indicated in Table~\ref{tab:dltV}.}
\end{figure}

In Fig.~\ref{Fig:udQM_Efrac} we present the energy per baryon, charge-to-mass ratio, and ratio of root-mean-square radius to baryon number for $ud$QM nuggets, which are obtained by taking large enough $m_{s0}$. The dashed curves indicate the properties of strangelets corresponding to those in Figs.~\ref{Fig:EpA_Slet} and \ref{Fig:frac_Slet}, where $m_{s0} = 96.0$ MeV. It is evident that adopting the same $\delta_V$ as in strangelets well reproduces the energy per baryon and charge-to-mass ratio of $ud$QM nuggets with TFA, and both of which are decreasing with $A$. In comparison with strangelets, we find $ud$QM nuggets become more massive with the energy per baryon increased by up to 40 MeV, while the magnitude is proportional to $f_S$ for the strangelets indicated in Fig.~\ref{Fig:frac_Slet}. For the cases obtained with the parameter set $C=0.7$ and $\sqrt{D}=129$ MeV, the properties of $ud$QM nuggets are similar to those of strangelets with rather small $f_S$. The energy per baryon of $ud$QM nuggets considered here are larger than that of the heaviest $\beta$-stable nucleus $^{266}$Hs with $M/A= 931.74$ MeV~\cite{Audi2017_CPC41-030001, Huang2017_CPC41-030002, Wang2017_CPC41-030003}, which permits the existence of $ud$QM nuggets as $^{266}$Hs and lighter nuclei do not decay into them. Nevertheless, only the $ud$QM nuggets obtained with $C=0.7$ and $\sqrt{D}=129$ MeV become stable as baryon number increases. In general, we note $ud$QM nuggets are more positively charged and less compact in comparison with strangelets. The charge-to-mass ratios obtained with various parameter sets are close to each other, which is mainly caused by the similar values of symmetry energy ($E_\mathrm{sym}\approx 15$ MeV) predicted by those parameters.

\begin{figure}
\includegraphics[width=\linewidth]{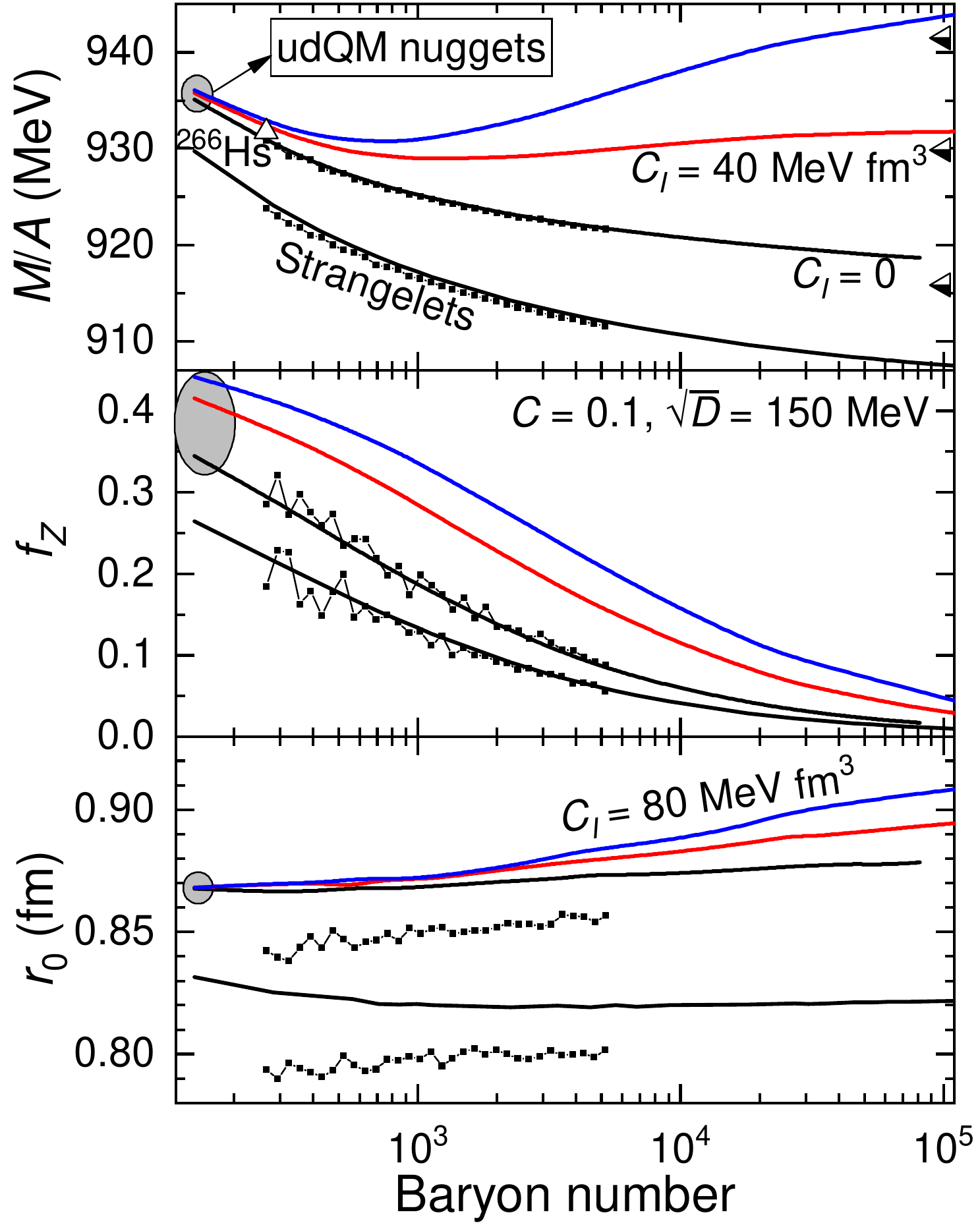}
\caption{\label{Fig:Nug_C01D150} Same as Fig.~\ref{Fig:udQM_Efrac} but adopting the parameter set $C=0.1$,  $\sqrt{D}=150$ MeV and various values of $C_I$. The curves encircled by the shaded areas indicate the properties of $ud$QM nuggets. The coefficient of the density derivative term is taken as $\delta_V=-1.5\times10^{-10}$ MeV$^{4}$, which is indicated in Table~\ref{tab:dltV}.}
\end{figure}

The symmetry energy of quark matter in Fig.~\ref{Fig:udQM_Efrac} mainly stems from the kinetic contributions, while in principle the interaction among quarks could alter its value~\cite{Chu2014_ApJ780-135, Jeong2016_NPA945-21, Chen2017_NPR34-20, Chu2019_PRC99-035802, Wu2019_AIPCP2127-020032}, e.g., the formation of $u$-$d$ quark Cooper pairs (2SC phase)~\cite{Jeong2016_NPA945-21}. In such cases, as proposed in our {\color{red}previous} study~\cite{Wang2021_Galaxies9-70}, we have included the third term in Eq.~(\ref{Eq:mI}) to examine the effects of quark matter symmetry energy $E_\mathrm{sym}$, which increases with $C_I$. As indicated in Fig.~\ref{Fig:Nug_C01D150}, we first fix $\delta_V$ ($=-1.5\times10^{-10}$ MeV$^{4}$) by reproducing the properties of $ud$QM nuggets and strangelets in the framework of TFA, where the parameter set $C=0.1$, $\sqrt{D}=150$ MeV, and $C_I=0$ is adopted. Since the third term in Eq.~(\ref{Eq:mI}) is proportional to the baryon number density $n_\mathrm{b}$, the corresponding contribution is thus insignificant in the surface region in comparison with the first term (${D}{n_\mathrm{b}^{-1/3}}$) as it provides a confining potential. In such cases, we expect varying $C_I$ has little impact on the values of $\delta_V$, where the value of $\delta_V$ fixed at $C_I=0$ is adopted for the cases with nonzero $C_I$.

Similar to the cases in Fig.~\ref{Fig:udQM_Efrac}, it is found that $ud$QM nuggets indicated in Fig.~\ref{Fig:Nug_C01D150} are more massive and positively charged than strangelets, while the radii become larger as well. We note that the $ud$QM nugget at $A=266$ is more stable than $^{266}$Hs adopting $C_I=0$, which is forbidden since no incidents of finite nuclei decaying into $ud$QM nuggets have been observed. This problem can be fixed if large symmetry energies with $C_I = 40$ and 80 MeV fm$^3$ are adopted, where $ud$QM nuggets become unstable comparing with finite nuclei. With the increase of $E_\mathrm{sym}$, there exist $ud$QM nuggets at $A\approx 1000$ that are more stable than others, where the energy per baryon is even smaller than the infinite quark matter indicated by the half-solid triangles in Fig.~\ref{Fig:Nug_C01D150}. In such cases, large $ud$QM nuggets will likely decay via fission and the surface of an $ud$QM star will fragment into a crust made of the $ud$QM nuggets and electrons, which resembles the cases of strange stars' crusts~\cite{Jaikumar2006_PRL96-041101, Alford2008_PRC78-045802}. As $E_\mathrm{sym}$ increases with $C_I$, in contrast to the cases predicting similar $f_Z$'s with $C_I=0$ in Fig.~\ref{Fig:udQM_Efrac}, the charge-to-mass ratio of $ud$QM nuggets increases with $C_I$. Due to the additional repulsive interactions from symmetry energy and Coulomb repulsion with larger $f_Z$, the ratio of root-mean-square radius to baryon number $r_0$ for $ud$QM nuggets becomes larger and increases with $A$, while that of strangelets remains almost constant.

\subsection{Compact stars}
If either SQM or $ud$QM becomes more stable than nuclear matter, there should exist stable quark stars comprised of those matter, i.e., strange stars~\cite{Itoh1970_PTP44-291, Alcock1986_ApJ310-261, Haensel1986_AA160-121} or $ud$QM stars~\cite{Zhao2019_PRD100-043018, Zhang2020_PRD101-043003, Cao2020, Zhang2021_PRD103-063018, Yuan2022}. For most of the cases considered here, we find that both strangelets and $ud$QM nuggets become more stable as baryon number $A$ increases. Then a stable quark star is bare on its surface, which is comprised of a sharp quark-vacuum interface covered by an electron cloud about 1 {\AA} thick. In addition to that, as indicated in Fig.~\ref{Fig:Nug_C01D150}, there exists $ud$QM nuggets at certain sizes ($A\approx 1000$) that are more stable than others if a large symmetry energy is employed, which leads to the formation of stable quark star crusts~\cite{Wang2021_Galaxies9-70}. According to various investigations on strange stars, it was shown that crusts made of strangelets and electrons could exist if the surface tensions $\sigma$ of SQM is small enough~\cite{Jaikumar2006_PRL96-041101, Alford2008_PRC78-045802}. Similar structures are thus expected to be formed on an $ud$QM star's surface~\cite{Wang2021_Galaxies9-70}.

Now let us examine such possibilities for $ud$QM stars in the framework of TFA, where the microscopic structures of their crusts are determined within a WS cell, i.e., adopting the single nucleus approximation. As an example, we take the parameter set $C=0.1$, $\sqrt{D}=150$ MeV, and $C_I=50$ MeV fm$^3$, so that $ud$QM nuggets at $A\approx 1000$ are more stable than others with $M/A<930$ MeV. Based on the numerical recipe introduced in Sec.~\ref{sec:the_TFA}, at fixed average baryon number density $n_\mathrm{b}$, we search for the most stable configuration among six types of structures (droplets, rods, slabs, tubes, bubbles, and uniform) with optimum WS cell sizes $R_\mathrm{W}$. For the nonuniform structures considered here, it is found that the droplet phase is the most stable configuration, which is similar to strange stars' crusts according to previous investigations~\cite{Alford2008_PRC78-045802}.

\begin{figure}[htbp]
\includegraphics[width=\linewidth]{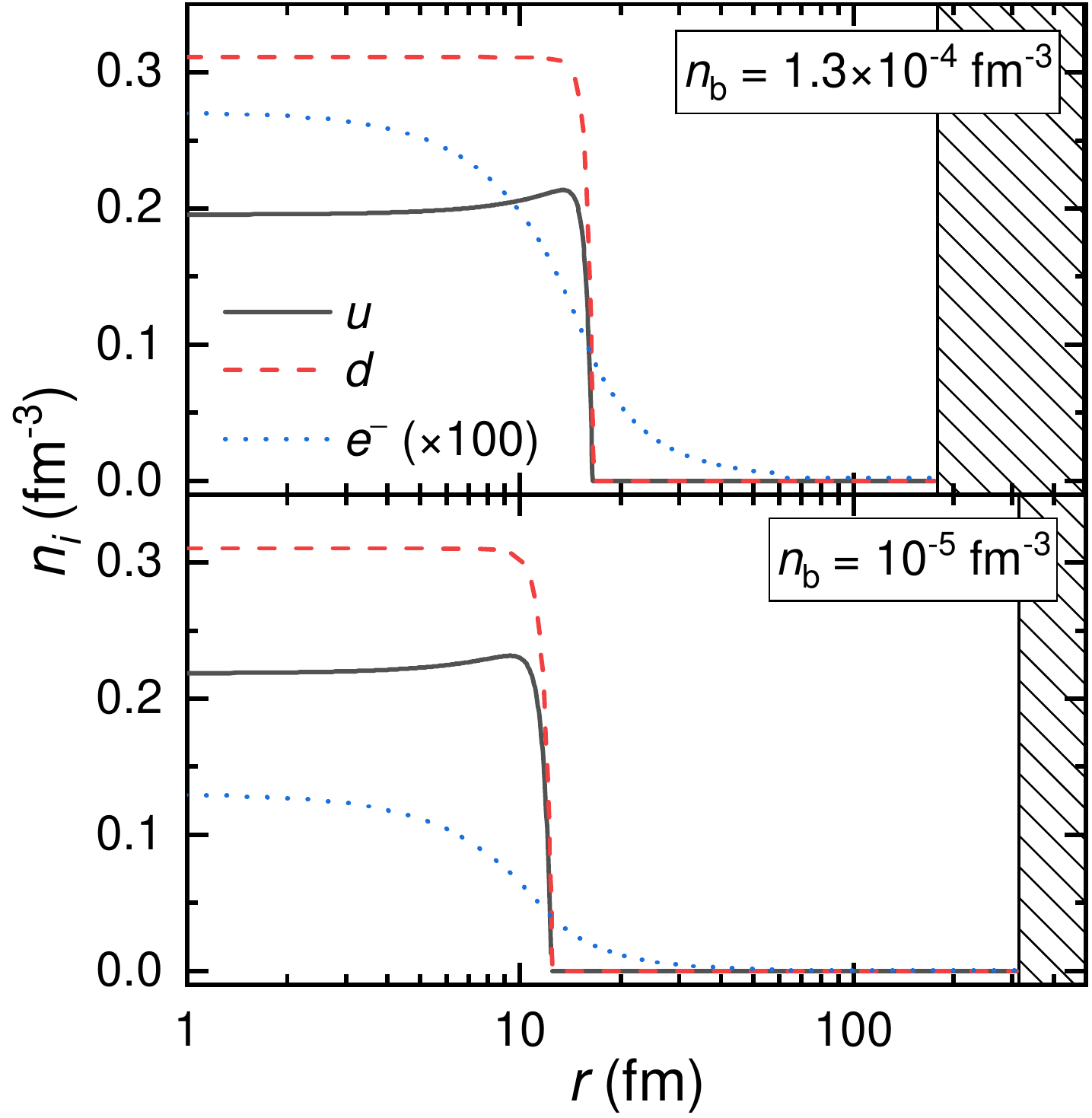}
\caption{\label{Fig:Dens} Density profiles of quarks and electrons in WS cells for droplet phase at $n_\mathrm{b}=1.3\times10^{-4}$ and $10^{-5}$ fm$^{-3}$. The boundary of the WS cell is indicated by the shaded region.}
\end{figure}

In Fig.~\ref{Fig:Dens} we present the density profiles of the droplet phase at $n_\mathrm{b}=1.3\times10^{-4}$ and $10^{-5}$ fm$^{-3}$, which is more stable than other nonuniform structures with the energy per baryon reduced by $\gtrsim 1$ MeV. The densities of quarks remain almost constant within the droplet and vanish at $r>R$ with $R$ (= 16.5 fm and 12.5 fm) being the position of quark-vacuum interface. As we increase $r$, the density of $u$ quarks increases and that of $d$ quarks decreases at $r\gtrsim R-10$ fm due to Coulomb repulsion, while both of which start to decrease at $r\approx R-3$ fm under the impacts of the density derivative term in Eq.~(\ref{eq:Lgrg_all}) and the confining potential. We have considered the charge screening effects with the electrons move freely and fulfill the constancy of chemical potential in Eq.~(\ref{eq:chem_cons}). It is found that $R$ is much smaller than the optimum WS cell radius $R_\mathrm{W}$ so that most of the space is occupied by electrons. Since the obtained $R_\mathrm{W}$ is rather large, as was done for the outer crusts of neutron stars~\cite{Xia2022_PRC105-045803}, we have divided the WS cell into two parts at $r=R_\mathrm{in}=60$ fm, where the electron density remain constant at $r>R_\mathrm{in}$.

\begin{figure}[!ht]
  \centering
  \includegraphics[width=\linewidth]{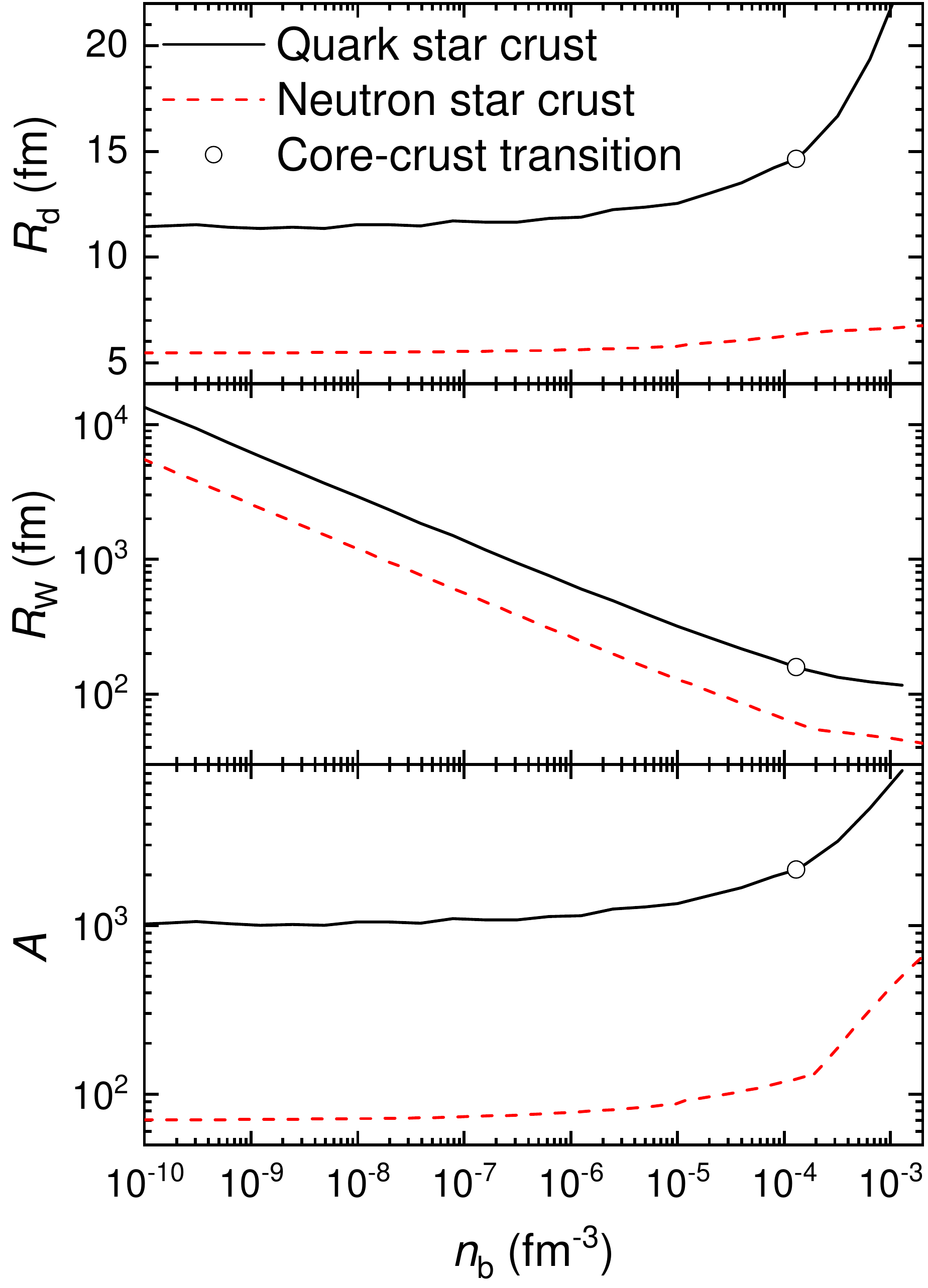}
  \caption{\label{Fig:Micro} Droplet size $R_\mathrm{d}$, WS cell radius $R_\mathrm{W}$, and baryon number $A$ for the droplet phase in $ud$QM stars' crusts. The corresponding structures in neutron stars' crusts are indicated with red dashed curves for comparison~\cite{Xia2022_PRC105-045803}.}
\end{figure}

Based on the density profiles illustrated in Fig.~\ref{Fig:Dens}, we can fix the droplet size $R_\mathrm{d}$ with
\begin{equation}
   R_\mathrm{d} = R_\mathrm{W}\left(\frac{9 n_\mathrm{b}^2}{\langle (n_u+n_d)^2 \rangle}\right)^{1/3},  \label{Eq:Rd}
\end{equation}
where $\langle (n_u+n_d)^2 \rangle = \int \left[n_u(\vec{r})+n_d(\vec{r})\right]^2 \mbox{d}^3 r/V$ with $V=4\pi R_\mathrm{W}^3/3$ being the WS cell volume. The corresponding quantities that characterize the microscopic structures of matter inside quark stars' crusts are illustrated in Fig.~\ref{Fig:Micro}, where the
droplet size $R_\mathrm{d}$, WS cell radius $R_\mathrm{W}$, and baryon number $A$ are presented as functions of baryon number density. It is found that inside the crusts of $ud$QM stars there typically exist $ud$QM nuggets with $R_\mathrm{d}\approx 12$ fm that arrange themselves in lattice structures, corresponding to the most stable $ud$QM nuggets at $A\approx 1000$ as indicated in Fig.~\ref{Fig:Nug_C01D150}. As the average baryon number density $n_\mathrm{b}$ increases, the size of $ud$QM nuggets increases and eventually the nonuniform structures becomes unstable in comparison with the uniform one, indicating a core-crust transition density for $ud$QM stars. Meanwhile, the WS cell radius $R_\mathrm{W}$ is much larger than the droplet size $R_\mathrm{d}$, which decreases as density increases. In Fig.~\ref{Fig:Micro} we have also included the microscopic structures of nuclear pasta in neutron stars, which shows the similar density-dependent behavior as those in $ud$QM stars. Nevertheless, the optimum droplet size, WS cell radius, and baryon number are much smaller, which is mainly attributed to the much smaller energy contributions from the surfaces of nuclei.

\begin{figure*}
\includegraphics[width=0.7\linewidth]{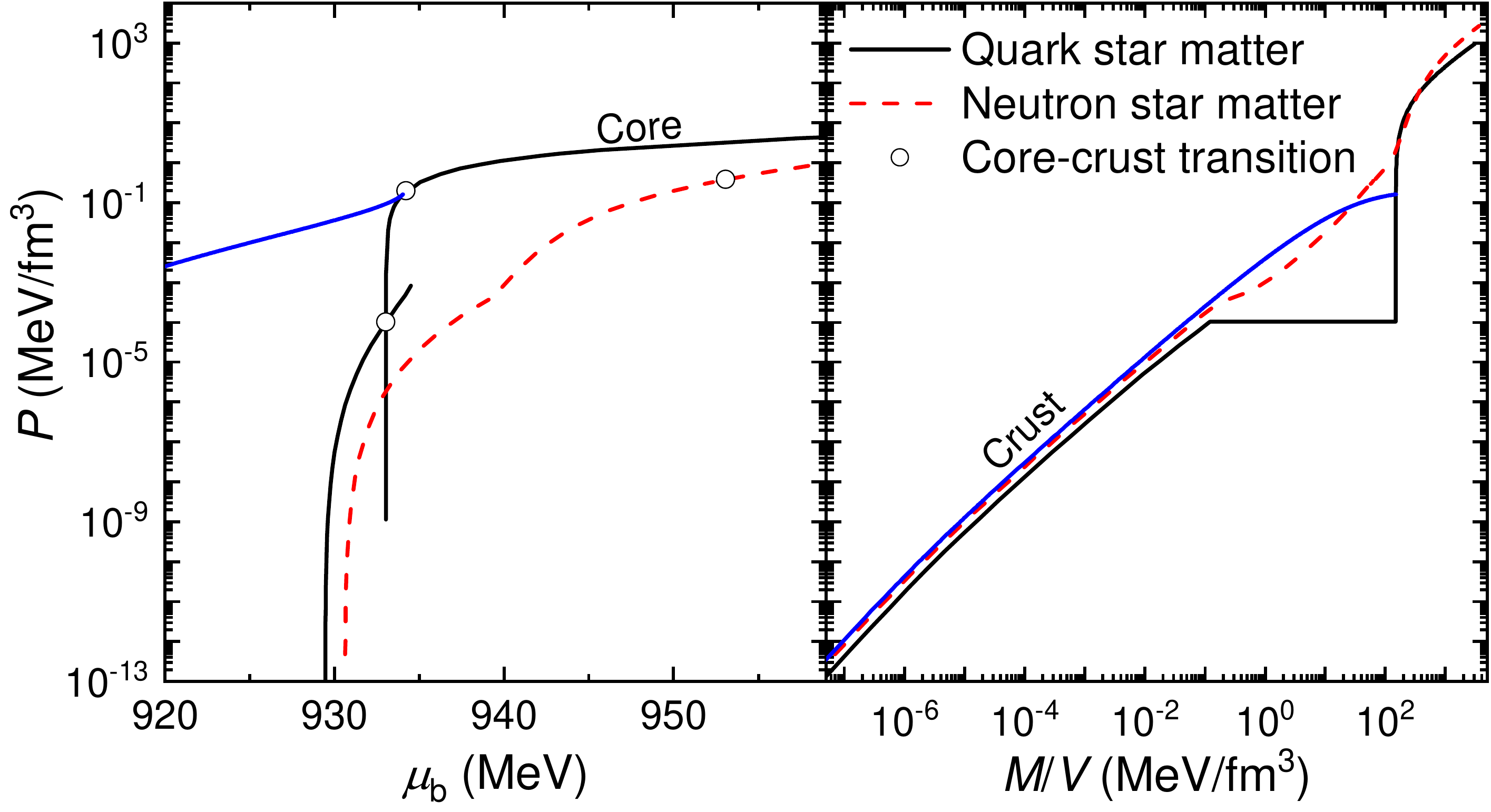}
\caption{\label{Fig:EOS} Pressures of dense stellar matter in $ud$QM stars and neutron stars as functions of baryon chemical potential $\mu_\mathrm{b}$ and energy density $M/V$. The open circles mark the critical pressure and chemical potential, below which the nonuniform structures take place.}
\end{figure*}

The equation of state (EOS) for dense stellar matter inside $ud$QM stars and neutron stars are then presented in Fig.~\ref{Fig:EOS}, where the corresponding microscopic structures are indicated in Fig.~\ref{Fig:Micro}. For the uniform matter in $ud$QM stars, the derivative terms in Eqs.~(\ref{Eq:Vi}-\ref{eq:mass}) vanish so that the corresponding properties of $ud$QM can be obtained easily by fulfilling the local charge neutrality condition $\sum_i q_in_i=0$ and $\beta$-equilibrium condition $\mu_e=\mu_\mu=\mu_d - \mu_u$. In our previous study~\cite{Wang2021_Galaxies9-70}, we have neglected the effects of charge screening and assumed vanishing surface tension, in which case the EOSs of the $ud$QM stars' crusts can be fixed based on Gibbs construction, i.e.,
\begin{equation}
    P_\mathrm{QM}(\mu_\mathrm{b},\mu_e) = 0. \label{Eq:phase_equiv1}
\end{equation}
Here $P_\mathrm{QM}$ represents the pressure of pure quark matter in the absence of leptons and $\mu_\mathrm{b}=\mu_u+2\mu_d$ the baryon chemical potential. The volume fraction $\chi$ of quark matter in $ud$QM star crusts is fixed by fulfilling the global charge neutrality condition
\begin{equation}
    \chi \left(\frac{2}{3}n_u-\frac{1}{3}n_d\right) = n_e. \label{Eq:ch0_equiv1}
\end{equation}
According to Eq.~(\ref{Eq:phase_equiv1}), the pressure is solely from electrons with $P=P_e(\mu_e)$, while the energy density is given by
\begin{equation}
  M/V = \chi M_\mathrm{QM}(\mu_\mathrm{b},\mu_e)/V + M_e(\mu_e)/V. \label{Eq:E_equiv1}
\end{equation}
The obtained results are then indicated by the blue solid curve with larger pressure $P$ in Fig.~\ref{Fig:EOS}, which corresponds to the limit of vanishing surface tension.

To fix the core-crust transition density indicated by the open circles in Fig.~\ref{Fig:Micro}, we consider the equilibrium condition between the uniform and nonuniform phases with
\begin{equation}
P^\mathrm{Uniform}(\mu_\mathrm{t}) = P^\mathrm{Nonuniform}(\mu_\mathrm{t})=P_\mathrm{t}. \label{Eq:phase_equiv}
\end{equation}
The EOS of $ud$QM stars is then fixed by combining the nonuniform one at $\mu_\mathrm{b}\leq \mu_\mathrm{t}$ and the uniform one at $\mu_\mathrm{b}> \mu_\mathrm{t}$, which is indicated by the black curves in Fig.~\ref{Fig:EOS}. We note that there is a large discrepancy between the electron chemical potentials of the two phases with $\mu_e^\mathrm{Uniform}(\mu_\mathrm{t}) > \mu_e^\mathrm{Nonuniform}(\mu_\mathrm{t})$. In such cases, similar to the cases of a strange star enveloped by a nuclear crust~\cite{Alcock1986_ApJ310-261}, there exists a gap of $\lesssim$1 {\AA} between the crust and core for an $ud$QM star, which is supported by Coulomb interaction. In principle, the phase equilibrium condition in Eq.~(\ref{Eq:phase_equiv}) is altered slightly under such circumstances with the local electron chemical potential vary smoothly from core to crust, e.g., similar to the cases illustrated in Refs.~\cite{Belvedere2012_NPA883-1, Rueda2014_PRC89-035804, Xia2016_SciBull61-172, Xia2016_PRD93-085025}. Nevertheless, due to the effects of charge screening, we expect such a variation in the core-crust transition regions is insignificant for the masses and radii of quark stars. Note that the gap vanishes if $\sigma=0$, i.e., those determined by Gibbs construction in Eqs.~(\ref{Eq:phase_equiv1}-\ref{Eq:E_equiv1}) with the same electron chemical potentials for the uniform and nonuniform phases at the core-crust transition regions in $ud$QM stars.

According to the left panel of Fig.~\ref{Fig:EOS}, it is evident that $ud$QM is more stable than nuclear matter with larger pressures. With neutrons dripping out of nuclei, there exists a liquid-gas mixed phase that forms the inner crusts of neutron stars, which is absent in $ud$QM stars with the crusts resemble the outer crusts of neutron stars. Due to the additional contributions of neutron gas, the corresponding core-crust transition pressure of dense stellar matter in neutron stars is thus larger than that of $ud$QM stars. The blue curves indicate the pressure of crustal matter in $ud$QM stars obtained with Gibbs construction, which correspond to the cases with vanishing surface tension. In comparison with the more realistic calculation adopting TFA, the results obtained with Gibbs construction should be viewed as an upper limit, which indicate a larger core-crust transition pressure predicted by Eq.~(\ref{Eq:phase_equiv1}) with $\chi=1$.

\begin{figure}
\includegraphics[width=\linewidth]{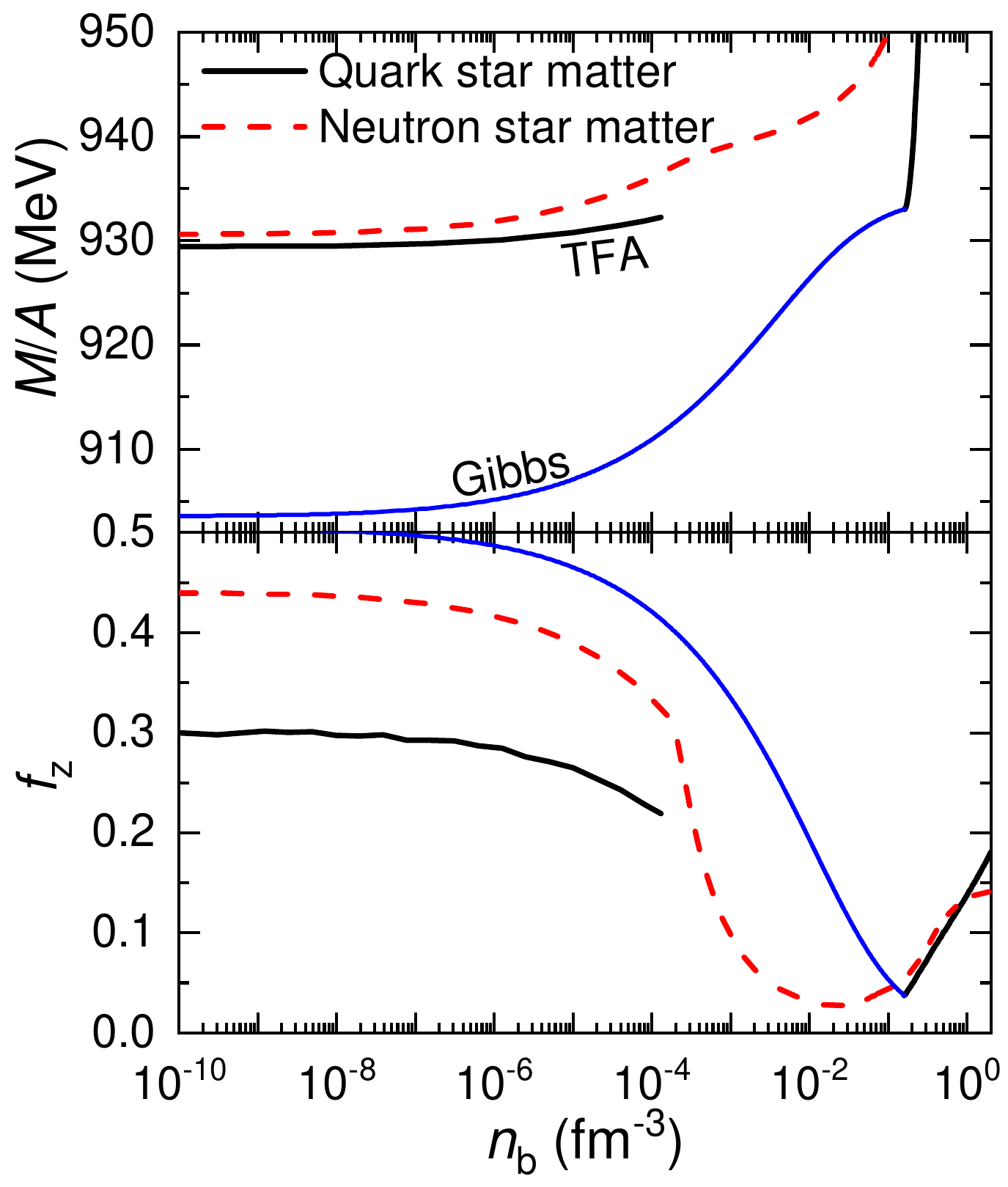}
\caption{\label{Fig:EpAfz} Energy per baryon and charge-to-mass ratio for dense stellar matter in $ud$QM stars and neutron stars as functions of baryon number density.}
\end{figure}

The EOSs for the dense stellar matter in $ud$QM stars and neutron stars are then presented in the right panel of Fig.~\ref{Fig:EOS}. The electrons play a major role for the EOSs of $ud$QM stars' crusts as well as neutron stars' outer crusts, which  exhibit similar trends at $M/V \lesssim 0.1$ MeV/fm$^3$. As indicated in Fig.~\ref{Fig:EpAfz}, for those obtained with Gibbs construction, the charge-to-mass ratio of $ud$QM nuggets in $ud$QM stars are lager than that of nuclei in neutron stars, where the situation reverse if we consider the interface effects with a density derivative term in Eq.~(\ref{eq:Lgrg_der}). Consequently, as the pressure increases with electron density, the pressure of the crustal matter in $ud$QM stars at $\sigma=0$ is larger than neutron stars and smaller if the more realistic calculation with TFA is adopted. A density jump is observed at the core-crust transition pressure in quark star matter, which becomes less evident if Gibbs construction is adopted. This is distinctively different from neutron star matter, where the transition takes place smoothly with the formation of liquid-gas mixed phase, i.e., nuclear pastas.

In Fig.~\ref{Fig:EpAfz} we present the energy per baryon and charge-to-mass ratio for dense stellar matter corresponding to the EOSs indicated in Fig.~\ref{Fig:EOS}. The energy per baryon of quark star matter is smaller than that of neutron star matter, which is permitted by current constraints of nuclear physics since finite nuclei do not decay into $ud$QM nuggets as illustrated in Fig.~\ref{Fig:udQM_Efrac}. We note there is a large density jump for the quark star matter obtained with TFA, which decreases if we reduce the energy contribution from the quark-vacuum interface and vanishes at $\sigma=0$ as indicated by the blue solid curve. In general, the obtained energy per baryon increases with density, while the energy per baryon of $ud$QM becomes smaller if the contributions from quark-vacuum interface is neglected. The obtained charge-to-mass ratio $f_Z$ for $ud$QM nuggets are smaller than nuclei in compact stars. Nevertheless, if the interface effects are neglected, $ud$QM nuggets become infinitesimal and $f_Z$ increases to 0.5 as the energy contribution from Coulomb interaction vanishes.

\begin{figure}
\includegraphics[width=\linewidth]{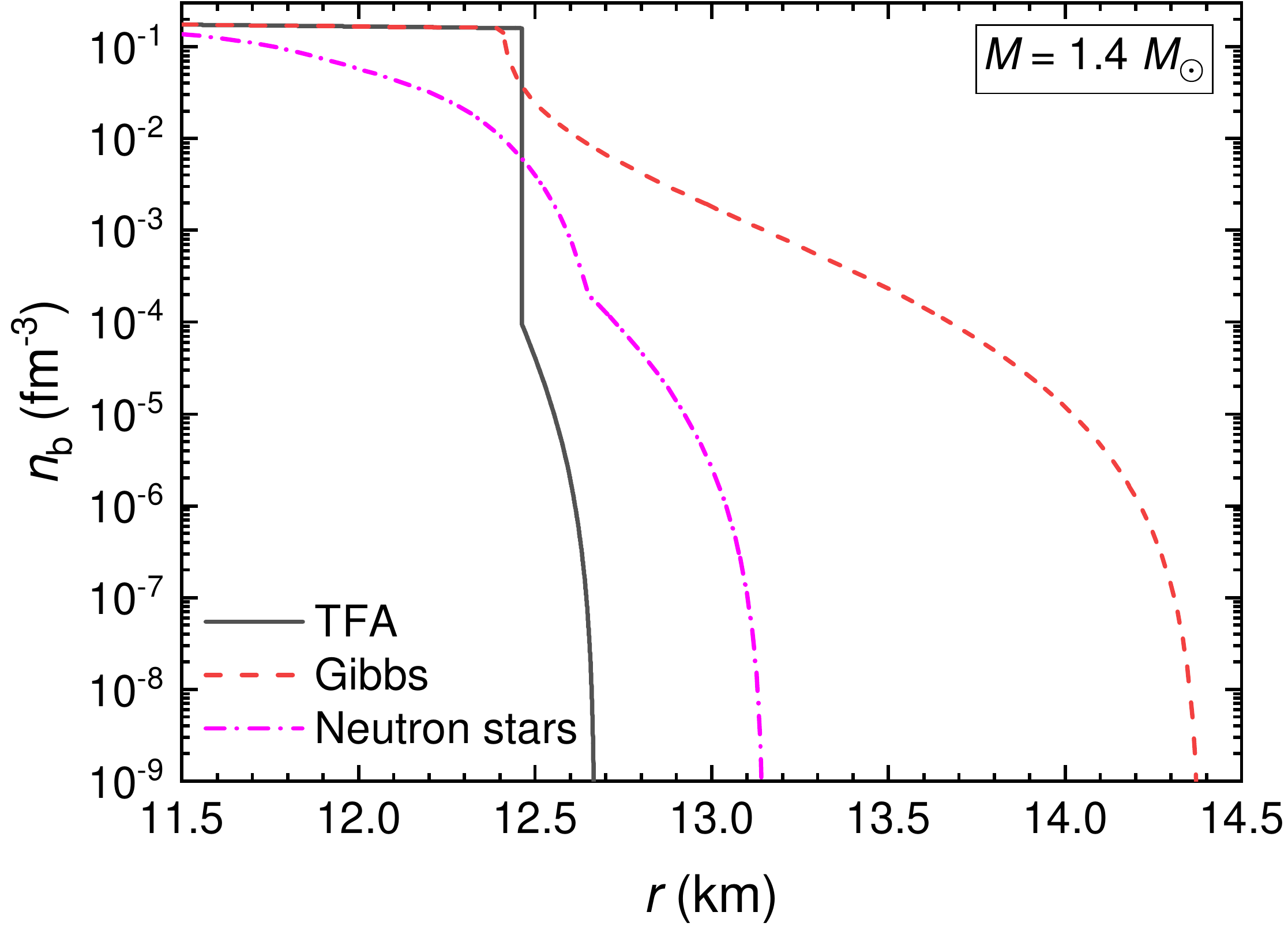}
\caption{\label{Fig:Intprofile} Density profiles of 1.4-solar-mass $ud$QM stars and neutron star.}
\end{figure}

Based on the EOSs of dense stellar matter in Fig.~\ref{Fig:EOS}, the structures of $ud$QM stars and neutron stars can be fixed by solving the Tolman-Oppenheimer-Volkov (TOV) equation. Due to the peculiar properties of EOSs around the core-crust transition densities, the obtained structures of compact stars and in particular the crusts vary significantly under different circumstances. To show this explicitly, in Fig.~\ref{Fig:Intprofile} we present the density profiles of 1.4-solar-mass $ud$QM stars and neutron star, where for $ud$QM stars both the realistic case with TFA and the extreme case with $\sigma=0$ are illustrated. In contrast to neutron stars, the density drops abruptly around the core-crust transition regions, which is attributed to the absence of inner crusts in $ud$QM stars. The $\sim$200 m thick crust is supported by the strong electric field of the core with a gap of $\lesssim$1 {\AA} between them, which resembles the cases of a strange star enveloped by a nuclear crust~\cite{Alcock1986_ApJ310-261}. Nevertheless, it is worth mentioning that the crust is stable and in equilibrium with the core, where the baryon chemical potentials take the same value. The drop of density becomes less evident if the interface effects are neglected with the crust EOS fixed by Gibbs construction, which predicts much thicker crusts ($\sim$2 km) for $ud$QM stars. Note that a larger core-crust transition density is predicted by Gibbs construction, while that of the realistic case with TFA may be altered slightly depending on the electron chemical potential and charge screening effects in the transition region~\cite{Belvedere2012_NPA883-1, Rueda2014_PRC89-035804, Xia2016_SciBull61-172, Xia2016_PRD93-085025}.

\begin{figure*}
\centering
\includegraphics[width=0.47\linewidth]{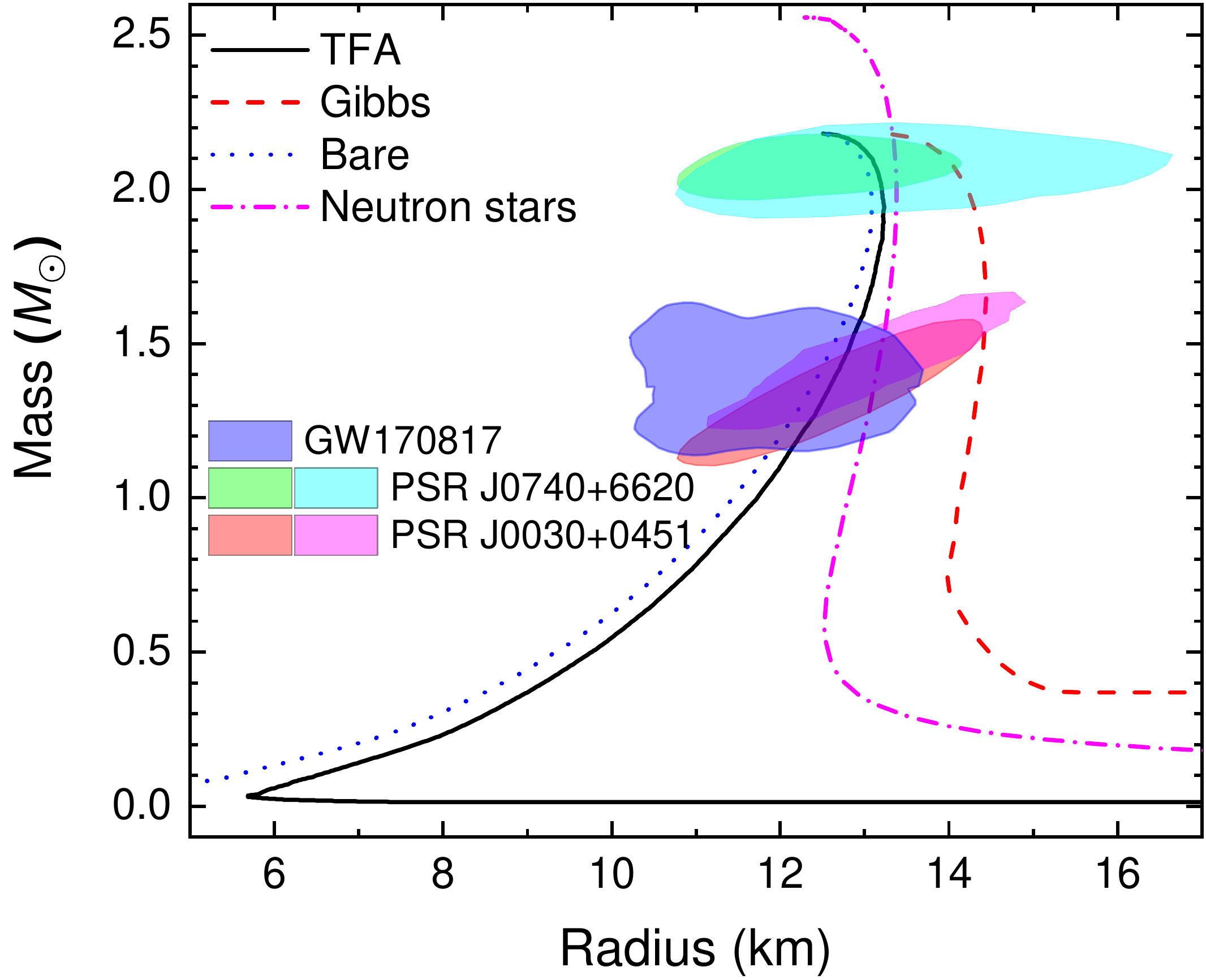}
\includegraphics[width=0.47\linewidth]{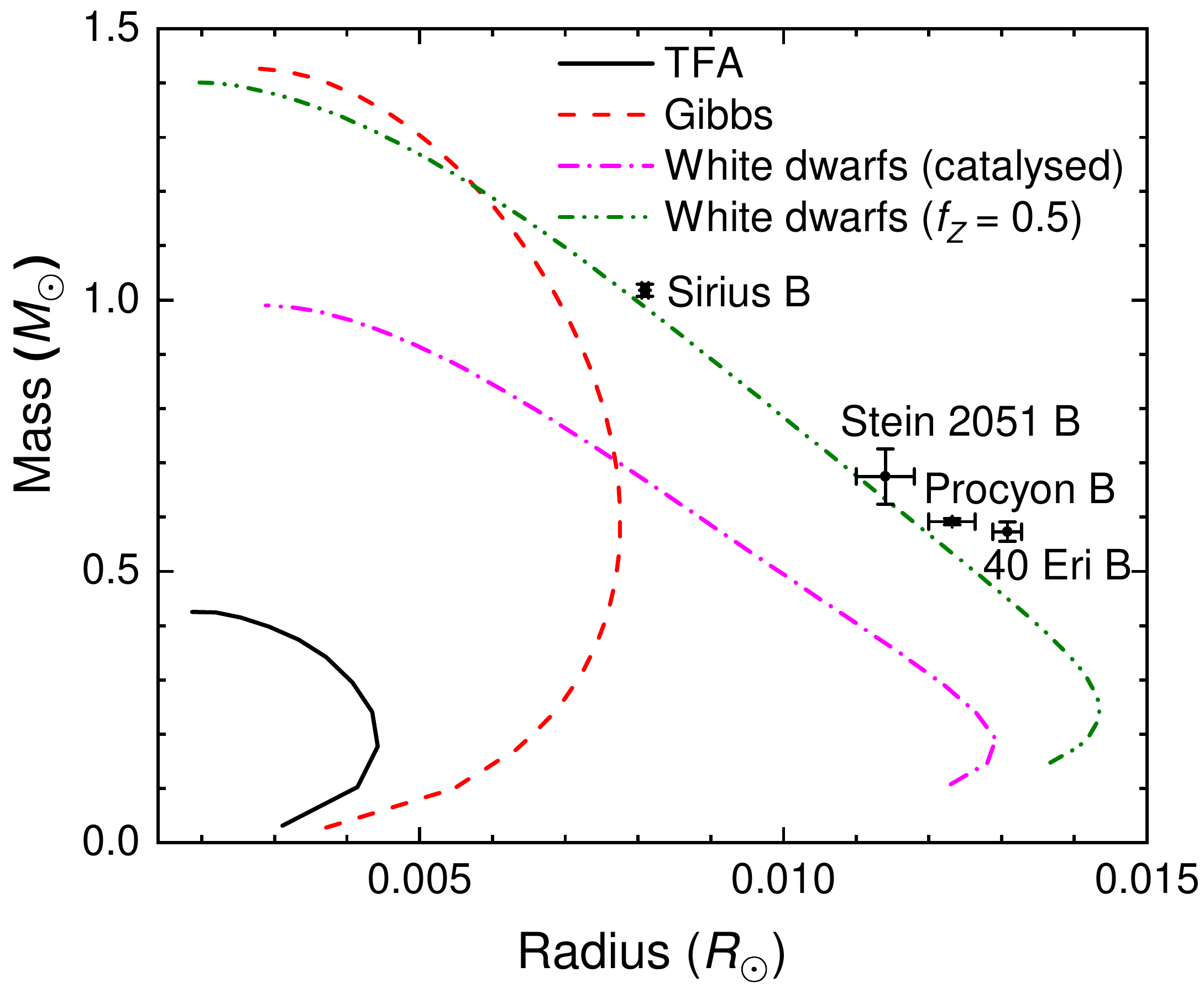}
\caption{\label{Fig:MR} Mass-radius relations of cold $ud$QM stars (left), neutron stars (left), $ud$QM dwarfs (right), and white dwarfs (right). The shaded areas indicate the constraints from the binary neutron star merger event GW170817~\cite{LVC2018_PRL121-161101} and the pulse profile modeling for PSR J0030+0451 and PSR J0740+6620~\cite{Riley2019_ApJ887-L21, Riley2021_ApJ918-L27, Miller2019_ApJ887-L24, Miller2021_ApJ918-L28}. The four dots in the right panel are the visual binaries Sirius B, Stein 2051 B, Procyon B, and 40 Eri B~\cite{Bond2017_ApJ848-16}.}
\end{figure*}

The mass-radius ($M$-$R$) relations of $ud$QM stars, neutron stars, $ud$QM dwarfs, and white dwarfs obtained under various circumstances are presented in Fig.~\ref{Fig:MR}, where the parameter set $C=0.1$, $\sqrt{D}=150$ MeV, and $C_I=50$ MeV fm$^3$ is employed. In the left panel, the $M$-$R$ relations for bare $ud$QM stars and $ud$QM stars with the crust EOSs fixed by TFA and Gibbs construction are presented. The $M$-$R$ relation of neutron stars obtained with the covariant density functional DD-LZ1~\cite{Wei2020_CPC44-074107, Xia2022_PRC105-045803} are indicated for comparison. The shaded regions show the constraints from the binary neutron star merger event GW170817 within 90\% credible region~\cite{LVC2018_PRL121-161101} and the pulse profile modeling for PSR J0030+0451 and PSR J0740+6620 according to the NICER and XMM-Newton Data within 68\% credible region~\cite{Riley2019_ApJ887-L21, Riley2021_ApJ918-L27, Miller2019_ApJ887-L24, Miller2021_ApJ918-L28}. It is found that the $ud$QM stars are consistent with the recent constraints on the masses and radii of pulsars, while the radii become slightly too large if $\sigma\rightarrow 0$. Nevertheless, we should mention that the tidal deformability of those $ud$QM stars ($\Lambda_{1.4}\approx 1050$) are too large according to the binary neutron star merger event GW170817~\cite{LVC2018_PRL121-161101}, which may be fixed adopting different parameters. The bare (unstable) $ud$QM stars are obtained with the EOS of uniform matter in Fig.~\ref{Fig:EOS}, where the radii is fixed at $P=0$. In such cases, the differences of radii between $ud$QM stars with and without crusts give roughly the crust thickness, since both the phase equilibrium conditions in Eqs.~(\ref{Eq:phase_equiv1}) and (\ref{Eq:phase_equiv}) predict core-crust transition pressure $P_\mathrm{t}\approx0$. It is found that the crust thickness is typically on the order of 200 m for $ud$QM stars with $M\gtrsim 0.03 M_\odot$. The crust thickness grows significantly if the energy contribution of quark-vacuum interface is neglected, which reaches a few kilometers for typical $ud$QM stars. In order to distinguish $ud$QM stars from neutron stars, we note that for realistic cases the radius of $ud$QM stars become rather small as the mass decreases, which can be identified if there exist compact objects with ultra small masses and radii, e.g., 4U 1746-37~\cite{Li2015_ApJ798-56}.

In the right panel of Fig.~\ref{Fig:MR}, we present the $M$-$R$ relations of $ud$QM dwarfs and white dwarfs adopting the same parameter sets as in the left panel. Two types of white dwarfs are examined, i.e., the catalyzed one that fulfills $\beta$-stability condition and the one with $f_Z=0.5$ that comprised of equal numbers of protons and neutrons. The observational masses and radii for white dwarfs in nearby visual binaries Sirius B, Stein 2051 B, Procyon B, and 40 Eri B are indicated with solid dots~\cite{Bond2017_ApJ848-16}, which are slightly larger than our predictions with $f_Z=0.5$. Nevertheless, by considering the effects of finite temperature and adopting larger $f_Z$, the masses and radii of white dwarfs should increase.  The masses of $ud$QM dwarfs are found to be larger than that of strangelet dwarfs~\cite{Alford2012_JPG39-065201}, which is mainly due to the larger charge-to-mass ratio of $ud$QM nuggets. The maximum mass of $ud$QM dwarfs increases and approaches to the Chandrasekhar limit if the interface effects are neglected, where the crustal EOS is fixed by Gibbs construction. As indicated in Fig.~\ref{Fig:MR}, the masses and radii of $ud$QM dwarfs obtained at $\sigma\rightarrow 0$ overlaps with those of massive white dwarfs. In such cases, the possible observation of compact stars with masses and radii smaller than traditional white dwarfs could be used to identify $ud$QM dwarfs~\cite{Kurban2022_PLB-137204}.

\section{\label{sec:con}Conclusion}
In this work we examine the interface effects on various types of quark matter objects, i.e., strangelets, $ud$QM nuggets, $ud$QM stars, and $ud$QM dwarfs. Particularly, we show that the interface effects can be described by introducing a density derivative term to the Lagrangian density and adopting Thomas-Fermi approximation. By adjusting the coupling constant $\delta_V$ for the density derivative term, the properties of strangelets and $ud$QM nuggets obtained by solving Dirac equations are then reproduced with Thomas-Fermi approximation. It is found that adopting certain parameter sets, $ud$QM nuggets at $A\lesssim300$ are unstable against decaying into nuclei but become stable at large baryon numbers, which is consistent with current nuclear physical constraints~\cite{Holdom2018_PRL120-222001}. Additionally, if we consider the cases with larger symmetry energies of quark matter, there exist $ud$QM nuggets at $A\approx 1000$ that are more stable than others. In such cases, large $ud$QM nuggets will decay via fission and the surface of an $ud$QM star will fragment into a crust made of the $ud$QM nuggets and electrons, which resembles the cases of strange stars' crusts~\cite{Jaikumar2006_PRL96-041101, Alford2008_PRC78-045802}. We then examine the microscopic structures of dense stellar matter in $ud$QM stars' crusts, where spherical and cylindrical approximations for the Wigner-Seitz cell are adopted~\cite{Pethick1998_PLB427-7, Oyamatsu1993_NPA561-431, Maruyama2005_PRC72-015802, Togashi2017_NPA961-78, Shen2011_ApJ197-20}. It is found that the droplet phase is the most stable configuration, which coincide with the strange stars' crusts according to previous investigations~\cite{Alford2008_PRC78-045802}. In such cases, the crusts in the surface regions of $ud$QM stars are made of electrons and $ud$QM nuggets in spherical shape ($A\approx 1000$). We then investigate the corresponding structures of $ud$QM stars and $ud$QM dwarfs. It is found that the crust thickness is typically $\sim$200 m for $ud$QM stars with $M\gtrsim 0.03 M_\odot$, which grows significantly and reaches a few kilometers if the energy contribution of quark-vacuum interface is neglected. For $ud$QM dwarfs, the masses and radii are smaller than traditional white dwarfs, while the maximum mass would increase and approach to the Chandrasekhar limit if the interface effects are neglected.

Further investigation on the effects of color superconductivity~\cite{Alford2001_PRD64-074017, Madsen2001_PRL87-172003, Oertel2008_PRD77-074015} and finite temperature~\cite{Ke2014_PRD89-074041, Gao2016_PRD94-094030} are necessary, which are expected to affect the energy contribution from quark-vacuum interface and alter the coupling constant $\delta_V$ for the density derivative term. In addition to Coulomb interaction, there may be strong attractive interactions among quark clusters, e.g., H-dibaryons~\cite{Sakai1997_NPA625-192, Glendenning1998_PRC58-1298, Lai2013_MNRAS431-3282}, which could lead to the very interesting conclusions of strangeon matter and strangeon stars~\cite{Xu2003_ApJ596-L59, Wang2017_ApJ837-81, Miao2022_IJMPE0-2250037} and should be examined in our future study.

\section*{ACKNOWLEDGMENTS}
This work was supported by National SKA Program of China No.~2020SKA0120300 and National Natural Science Foundation of China (Grant Nos.~11875052, 12005005  \& 11673002).




%

\end{document}